\documentclass[12pt,draftclsnofoot,onecolumn]{IEEEtran}

\usepackage{amsmath,graphics,amssymb,epsfig,subfigure,color,cite}
\usepackage{array}
\usepackage{multirow}
\usepackage{enumerate}
\usepackage{algorithm}
\usepackage{algpseudocode}
\usepackage{algcompatible}
\usepackage{verbatim}

\algblockdefx{FORP}{ENDFORP}[1]%
  {\textbf{for }#1 \textbf{do in parallel}}%
  {\textbf{end}}
\algblockdefx{FOR}{ENDFOR}[1]%
  {\textbf{for }#1 \textbf{do}}%
  {\textbf{end}}
\algblockdefx{IF}{ENDIF}[1]%
  {\textbf{if }#1 \textbf{then}}%
  {\textbf{end}}
\algnewcommand\algorithmicprocedure{\textbf{function}}
\algnewcommand\FUNC{\item[\algorithmicprocedure]}%
\algnewcommand\algorithmicendprocedure{\textbf{end function}}
\algnewcommand\ENDFUNC{\item[\algorithmicendprocedure]}%
\usepackage{bm}
\usepackage{float}
\usepackage{makecell}

\usepackage{lipsum}
\usepackage{capt-of}

\floatname{algorithm}{Procedure}

\newcommand{\argmax}{\operatornamewithlimits{argmax}}
\newcommand{\argmin}{\operatornamewithlimits{argmin}}
\makeatletter
\newcommand{\vast}{\bBigg@{4.5}}
\newcommand{\Vast}{\bBigg@{7.5}}
\makeatother

\begin{document}
    \title{MIMO Detection under Hardware Impairments: Learning with Noisy Labels}
    
    \author{Jinman Kwon, Seunghyeon Jeon, Yo-Seb Jeon, \IEEEmembership{Member,~IEEE}, \\and H. Vincent Poor, \IEEEmembership{Life Fellow,~IEEE}
    \thanks{Jinman Kwon, Seunghyeon Jeon, Yo-Seb Jeon are with the Department of Electrical Engineering, POSTECH, Pohang, Gyeongbuk 37673, South Korea (e-mail: jinman.kwon@postech.ac.kr; seunghyeon.jeon@postech.ac.kr; yoseb.jeon@postech.ac.kr).}
    \thanks{H. Vincent Poor is with the Department of Electrical Engineering, Princeton University, Princeton, NJ 08544 (e-mail: poor@princeton.edu).}
    \thanks{This work was presented in part at the 2022 IEEE Global Communications Conference (GLOBECOM)  \cite{Conference}.}
    }
 
	\vspace{-2mm}
	
	\maketitle
	\vspace{-12mm}
 
\begin{abstract}  
    This paper considers a data detection problem in multiple-input multiple-output (MIMO) communication systems with hardware impairments.
    To address challenges posed by nonlinear and unknown distortion in received signals, two learning-based detection methods, referred to as model-driven and data-driven, are presented.
    The model-driven method employs a generalized Gaussian distortion model to approximate the conditional distribution of the distorted received signal.
    By using the outputs of coarse data detection as noisy training data, the model-driven method avoids the need for additional training overhead beyond traditional pilot overhead for channel estimation. 
    An expectation-maximization algorithm is devised to accurately learn the parameters of the distortion model from noisy training data.
    To resolve a model mismatch problem in the model-driven method, the data-driven method employs a deep neural network (DNN) for approximating a-posteriori probabilities for each received signal.
    This method uses the outputs of the model-driven method as noisy labels and therefore does not require extra training overhead.
    To avoid the overfitting problem caused by noisy labels, a robust DNN training algorithm is devised, which involves a warm-up period, sample selection, and loss correction. 
    Simulation results demonstrate that the two proposed methods outperform existing solutions with the same overhead under various hardware impairment scenarios. 
    \end{abstract}


\section{Introduction}\label{Sec:Intro}
	Multiple-input multiple-output (MIMO) communication in the millimeter wave (mmWave) or terahertz (THz) bands is a key technology for the current and next generations of wireless networks \cite{5G,6G}.
	From a theoretical perspective, it is possible to provide data rates beyond hundreds of Gbits/sec by utilizing both the spatial multiplexing available in MIMO systems and the extremely large bandwidths available in the mmWave/THz bands.  
	This technology, however, brings new challenges in implementing and designing ideal RF hardware components at both a transmitter and a receiver.   
	For example, high-resolution digital-to-analog converters (DACs) and analog-to-digital converters (ADCs) have a power consumption that linearly increases with a sampling rate; thereby, the power consumed by these components becomes prohibitive when operating over extremely large bandwidths \cite{lowADC : 3, lowADC : Magazine}.
	Another example is the power amplifier (PA), one of the key RF components at the transmitter; the saturated output power of the PA using current technology decreases as the carrier frequency approaches $100$ GHz \cite{HI : PA 1}.
	Further, the design of other RF components such as local oscillators and mixers also becomes more difficult as the carrier frequency increases \cite{HI:THz}.

    Recently, the use of non-ideal RF components has attracted growing attention as a low-cost and tangible solution to circumvent the hardware-oriented challenges in future MIMO communications \cite{HI:THz, HI : PA 2, HI : PA 3, HI : book, HI : mmWave}. 
    When employing non-ideal RF components, however, hardware impairments such as nonlinear saturation at the PA and quantization error at the ADCs are inevitable and therefore lead to nonlinear distortion in baseband received signals \cite{HI : book, HI : mmWave}. 
    To alleviate the distortion caused by non-ideal RF components at the transmitter, several compensation methods have been developed by leveraging a parametric model \cite{Learning(Model) : 1, Learning(Model) : 2, Learning(Model) : 3} or a deep neural network (DNN) \cite{Learning DNN : 3 (Predistortion), Learning DNN : 4 (Post compensation)}.
    The common idea of these methods is to approximate the nonlinear distortion by using a proper model, and then to learn the parameters of this model to compensate for its effects.
    Unfortunately, even if these compensation methods are applied, they do not guarantee perfect compensation or cancellation of the distortion. As a result, residual effects are still problematic in baseband processing at the receiver. Moreover, these methods can only address the distortion caused by certain RF components. This fact implies that baseband received signals are still subject to distortion caused by RF components that cannot be addressed by compensation methods.

    The presence of nonlinear distortion in baseband received signals poses several challenges in the design of MIMO channel estimation and data detection methods, which are the most important baseband processes at the receiver in MIMO systems.
    Most traditional channel estimation and data detection methods are developed based on a linear MIMO channel model with additive white Gaussian noise (AWGN).
    The performance of these methods, however, is severely degraded when the baseband receive signals are contaminated by the nonlinear distortion. 
    For example, for a received signal subjected to nonlinear distortion, MIMO channel estimation based on the least-squares method does not provide the maximum-likelihood (ML) optimal channel estimate, unlike in linear MIMO channels. 
    Moreover, in the presence of nonlinear distortion, MIMO data detection based on the minimum Euclidean distance criterion is no longer optimal in the sense of minimizing the detection error probability. 
    Theoretically, it would be possible to design optimal channel estimation and data detection methods if complete knowledge of the nonlinear distortion is available at the receiver.
    Unfortunately, the distortion caused by non-ideal RF components depends on hardware designs, specifications, and manufacturers \cite{HI : book, HI : mmWave}. 
    Therefore, it is not only difficult to obtain the complete knowledge of the nonlinear distortion, but it is also challenging to create a universal model that can accurately characterize the behavior of the distortion.

    Deep learning (DL) has a powerful ability to learn and optimize nonlinear systems, which has resulted in remarkable success across various fields and applications. This success has naturally led to the development of DL-based MIMO detectors that leverage training data to learn nonlinear relationships between transmitted and received signals \cite{DNN Application : 1 (DetNet), DNN Application : 2 (ViterbiNet), DNN Application : 3 (OAMPNet)}. 
    DL-based MIMO detection has also been studied for addressing nonlinear and unknown distortion caused by hardware impairments \cite{Learning DNN : 1 (Naive), Learning DNN : 2 (MP), Learning DNN : 5 (Lord-Net)}.
    Data-driven DL detectors based on black-box DNN architectures were introduced and analyzed in \cite{Learning DNN : 1 (Naive)}, which can be applied to MIMO systems with hardware impairments.
    Model-based DL detectors based on iterative algorithms were developed for MIMO systems with low-resolution ADCs \cite{Learning DNN : 5 (Lord-Net)}, or for MIMO systems with non-ideal PA and I/Q imbalance \cite{Learning DNN : 2 (MP)}.
    By incorporating domain knowledge into the design of a DNN architecture, these detectors require relatively low training overhead compared to detectors that use black-box DNN architectures.
    However, the model-based DL detectors still rely on some prior knowledge about hardware impairments including the existence of the hardware impairments and the structure of their input-output features.  
    Moreover, the training overhead required by these detectors is much larger than the pilot overhead required for traditional MIMO channel estimation and therefore may not be affordable in practical communication systems.

    MIMO detection methods based on machine learning approaches, other than DL, have also been studied to address the challenges posed by nonlinear and unknown distortion caused by hardware impairments \cite{Learning : 1 (SLD), Learning : 2 (SLD), Learning : 3 (ELM)}.
    A representative example is a supervised-learning approach in \cite{Learning : 1 (SLD), Learning : 2 (SLD)} which explicitly learns the empirical distribution of the distorted received signals from training data. 
    Although this approach was originally developed for MIMO systems with low-resolution ADCs, it can be extended to address the composite effect of various hardware impairments in MIMO systems. 
    As long as hardware impairments are time-invariant, supervised-learning-based methods in \cite{Learning : 1 (SLD), Learning : 2 (SLD)} are capable of approaching ML-optimal detection performance as the size of the training data increases. 
    These methods, however, require significant training overhead that increases exponentially with the number of transmit antennas.
    To reduce the training overhead for supervised learning, an extreme learning machine (ELM) approach was developed in \cite{Learning : 3 (ELM)} for MIMO systems with non-ideal PAs and low-resolution ADCs. 
    One significant advantage of this approach is that its training overhead can be significantly reduced compared to other DL or supervised learning approaches.
    Unfortunately, the ELM approach does not guarantee optimal detection performance, and may result in inferior performance when compared to ML detection.
    To the best of the authors' knowledge, maximizing MIMO detection performance under hardware impairments with practical training overhead still remains an open problem.

    In this paper, we consider the data detection problem in MIMO systems with hardware impairments. 
    For these systems, we propose two learning-based detection methods that can overcome the challenges posed by nonlinear and unknown distortion in baseband received signals.
    Our proposed methods utilize noisy training data generated from coarse data detection and therefore avoid the need for additional training overhead beyond traditional pilot overhead for channel estimation. 
    Our simulations demonstrate that the proposed detection methods achieve superior performance compared to other practical solutions with the same overhead.
    The major contributions of this paper are summarized as follows:
    \begin{itemize}
        \item 
        We propose a model-driven detection method for MIMO systems with hardware impairments.
        The main idea of this method is to leverage a parametric model to approximate the conditional distribution of the distorted received signal.
        In particular, we newly introduce a generalized Gaussian distortion model that generalizes the conventional additive distortion model in \cite{HI : PA 2}.
        We then utilize the outputs of coarse data detection for generating {\em noisy} training data to learn the parameters of our distortion model.
        Based on the expectation maximization (EM) principle in \cite{NLNN}, we develop a robust parameter learning algorithm which estimates not only the parameters of the distortion model, but also transition probabilities from noisy labels to true labels.

        \item 
        When there is a significant difference between the actual conditional distribution of the distorted received signal and the generalized Gaussian distortion model, the performance of our model-driven method may be compromised.
        To overcome this limitation, we propose a data-driven detection method based on DL, which does not rely on any assumptions regarding the distribution of the distorted received signals.
        The main idea of our data-driven method is to leverage a DNN to approximate a-posteriori probabilities (APPs) for each received signal. 
        To avoid the need for extra training overhead, we generate noisy training data by utilizing the transmitted symbol indexes estimated by the model-driven method. 
        For robust training of DNN parameters, we devise a robust DNN training algorithm which involves warm-up period, sample selection, and loss correction. 
        To this end, we integrate robust DNN training techniques from \cite{Deep : SELFIE} and \cite{Deep : Self-adaptive} in a judicious manner.

        \item 
        We evaluate the symbol-error-rate (SER) performance of our proposed detection methods against existing methods using simulations. In these simulations, we consider various hardware impairment scenarios, including a simple additive distortion scenario and a more realistic scenario that accounts for the effects of non-ideal PAs and low-resolution ADCs. Simulation results demonstrate that the proposed detection methods outperform existing ones for both additive and realistic scenarios. Our results also show that the proposed data-driven method achieves superior performance compared to the proposed model-driven method in the realistic scenario. Simulation results for time-varying channels demonstrate that the proposed methods exhibit robustness against nonlinear distortion as well as temporal channel variations.
        
    \end{itemize}

    This work builds on \cite{Conference}, in which we presented only the model-driven detection method for MIMO systems with hardware impairments. In this current paper, we have newly developed a data-driven detection method based on a DL approach to overcome the limitations of the model-driven method. Additionally, we have enhanced the simulation study by considering various hardware impairment scenarios including time-varying channels, and also by demonstrating the superiority of the data-driven method over the model-driven method. 

    The remainder of the paper is organized as follows. 
    In Sec.~\ref{Sec:SystemModel}, we present a MIMO communication system with hardware impairments and introduce a conventional additive distortion model. 
    In Sec.~\ref{Sec:ModelDriven}, we propose a model-driven detection method for MIMO systems with hardware impairments. 
    In Sec.~\ref{Sec:DataDriven}, we propose a data-driven detection method that overcomes the limitation of the model-based method.
    In Sec.~\ref{Sec:Simul}, we provide simulation results to verify the superiority of the proposed detection methods. 
    Finally, in Sec.~\ref{Sec:Conclusion}, we present our conclusions and future research directions.
    
    {\em Notation:} Lowercase, boldface lowercase, and boldface uppercase letters denote scalar, column vectors, and matrices, respectively.
    $(\cdot)^T$ is the transpose, $(\cdot)^H$ is the conjugate transpose, $\mathbb{E}[\cdot]$ is the expectation and $\mathbb{P}(\cdot)$ is the probability.
    For a scalar $a$, $\vert a \vert$ is the absolute value.
    For a vector ${\bf a}$, $\Vert {\bf a} \Vert$ is the Euclidean norm, $({\bf a})_i$ is the $i$-th element of ${\bf a}$.
    For a matrix ${\bf A}$, $({\bf A})_{i,j}$ is the element in the $i$-th row and $j$-th column.
    ${\bf I}_m$ is an $m\times m$ identity matrix, and ${\bf 0}_n$ is an $n$-dimensional all-zero vector.
    $\mathcal{CN}(\mu,{\bf C})$ represents the distribution of circularly-symmetric complex Gaussian random vector with mean vector $\mu$ and covariance matrix ${\bf C}$.
    ${\sf Re}\{\cdot\}$ and ${\sf Im}\{\cdot\}$ denote the real and imaginary parts of a complex-valued matrix, respectively.

\section{System Model}\label{Sec:SystemModel}
    In this section, we present a MIMO communication system with hardware impairments. We then introduce a conventional additive distortion model considered in \cite{HI : PA 2} along with its limitations. 
    
    \begin{figure}[t]
       \centering \vspace{-3mm}
       {\epsfig{file=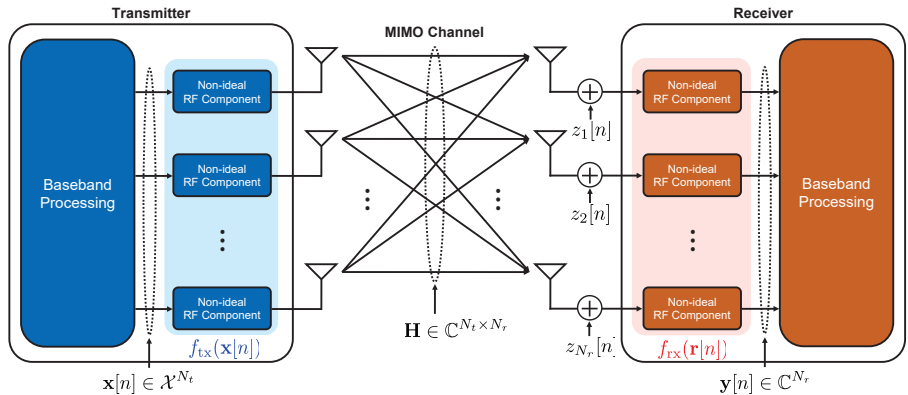, width = 12cm}}\vspace{-3mm}
       \caption{Illustration of a MIMO communication system under hardware impairments.} \vspace{-3mm}
       \label{fig:System}
    \end{figure}     
    
    \subsection{MIMO System with Hardware Impairments}
    We consider a frequency-flat MIMO communication system in which a transmitter equipped with $N_t$ antennas communicates with a receiver equipped with $N_r$ antennas.
    In particular, we assume that both the transmitter and the receiver are equipped with non-ideal RF components. 
    The MIMO system considered in our work is illustrated in Fig.~1.

    Let ${\bf x}[n] = \big[x_1[n],\cdots,x_{N_t}[n]\big]^{\sf T}\in\mathcal{X}^{N_t}$ be a data symbol vector sent at time slot $n\in\{1,\ldots,T\}$, where $T$ is the duration of data transmission.
    Each data symbol $x_i[n]$ is assumed to be drawn from a constellation set $\mathcal{X}$ with size $M=|\mathcal{X}|$ such that $\mathbb{E}[| x_i[n]|^2]=1$. 
    In the considered system, the data symbol vector is distorted before its transmission by the non-ideal RF components at the transmitter.
    Hence, a transmitted signal at time slot $n$ is expressed as ${\bf s}[n] = f_{\rm tx}({\bf x}[n])$, where $f_{\rm tx}:\mathbb{C}^{N_t}\rightarrow \mathbb{C}^{N_t}$ is a distortion function that represents the effect of the non-ideal RF components at the transmitter.  
    If the receiver is equipped with ideal RF components, an {\em ideal} received signal is expressed as
    \begin{align}
        {\bf r}[n] = {\bf H}f_{\rm tx}({\bf x}[n]) + {\bf z}[n], 
    \end{align}
    where ${\bf H}\in\mathbb{C}^{N_r\times N_t}$ is a MIMO channel matrix, and $\bold{z}[n]\in\mathbb{C}^{N_r}$ is additive white Gaussian noise (AWGN) distributed as ${\bf z}[n] \sim \mathcal{CN}({\bf 0}_{N_r},\sigma^2 {\bf I}_{N_r})$.
    However, in the considered system, due to the non-ideal RF components at the receiver, a baseband received signal at time slot $n$ is distorted as follows: 
    \begin{align}
        {\bf y}[n] = f_{\rm rx}\big({\bf r}[n] \big) = f_{\rm rx}\big({\bf H}f_{\rm tx}({\bf x}[n])+{\bf z}[n]\big),
    \end{align}
    where $f_{\rm rx}:\mathbb{C}^{N_r}\rightarrow \mathbb{C}^{N_r}$ is a distortion function that represents the effect of the non-ideal RF components at the receiver.  

    \subsection{Additive Distortion Model}\label{eq:AddModel}
    To capture the distortion caused by the non-ideal RF components at both the transmitter and the receiver, a simple additive distortion model was introduced in \cite{HI : PA 2}.
    In this prior work, the distortion function $f_{\rm tx}(\cdot)$ at the transmitter is modeled as
    \begin{align}
        f_{\rm tx}({\bf x}[n] )  = {\bf x}[n] + {\bm \eta}_{\rm tx}[n],
    \end{align}
    where ${\bm \eta}_{\rm tx}[n]\in \mathbb{C}^{N_t}$ is an additive distortion noise at the transmitter, which is independent with ${\bf x}[n]$ and distributed as ${\bm \eta}_{\rm tx}[n] \sim \mathcal{CN}\big({\bf 0}_{N_t},\kappa_{\rm tx}{\bf I}_{N_t}\big)$.
    Similarly, the distortion function $f_{\rm tx}(\cdot)$ at the receiver is modeled as
    \begin{align}
        f_{\rm rx}({\bf r}[n] )  = {\bf r}[n] + {\bm \eta}_{\rm rx}[n],
    \end{align}
    where ${\bm \eta}_{\rm rx}[n]\in \mathbb{C}^{N_r}$ is an additive distortion noise at the receiver, which is independent with ${\bf r}[n]$ and distributed as ${\bm \eta}_{\rm rx}[n] \sim \mathcal{CN}\big({\bf 0}_{N_r}, \kappa_{\rm rx}{\bf H}{\bf H}^{\sf H}\big)$.
    The assumption on the Gaussian distributions of ${\bf \eta}_{\rm tx}[n]$ and ${\bf \eta}_{\rm rx}[n]$ can be justified by the central limit theorem when these additive noises model the aggregate effect of many residual hardware impairments \cite{HI : PA 2}.
    Based on the above model, the distorted received signal at time slot $n$ can be rewritten as 
    \begin{align}\label{eq:additive_model}
        {\bf y}[n] 
        &= {\bf H}\big({\bf x}[n] + {\bf \eta}_{\rm tx}[n]\big) + {\bf \eta}_{\rm rx}[n] + {\bf z}[n] \nonumber \\
        &= {\bf H}{\bf x}[n] + {\bf z}_{\rm eff}[n],
    \end{align}
    where ${\bf z}_{\rm eff}[n] =  {\bf H}{\bf \eta}_{\rm tx}[n] + {\bf \eta}_{\rm rx}[n] + {\bf z}[n]$ is an effective noise at time slot $n$. 
    It is noticeable that the distribution of ${\bf z}_{\rm eff}[n]$ is given by 
    \begin{align}\label{eq:additive_noise}
        {\bf z}_{\rm eff}[n]  \sim \mathcal{CN}\big({\bf 0}_{N_r},  (\kappa_{\rm tx} + \kappa_{\rm rx}){\bf H}{\bf H}^{\sf H} + {\sigma}^2{\bf I}_{N_{r}} \big),
    \end{align}
    under the premise that ${\bf \eta}_{\rm tx}[n]$, ${\bf \eta}_{\rm rx}[n]$, and ${\bf z}[n]$ are statistically independent.

    A major limitation of the additive distortion model in \eqref{eq:additive_model} is that the effects of deterministic distortion are ignored in this model. 
    Two representative examples are PAs and ADCs which produce deterministic outputs for their inputs \cite{Learning : 3 (ELM), HI : PA 1}.
    Another limitation is that the exact values of $\kappa_{\rm tx}$ and $\kappa_{\rm rx}$ are unknown in general because these values depend on hardware designs, specifications, and manufacturers that can significantly differ across communication devices \cite{HI : book, HI : mmWave}. 

\section{Proposed Model-Driven MIMO Detection}\label{Sec:ModelDriven}
\begin{figure}[t]
      \centering \vspace{-3mm}
      {\epsfig{file=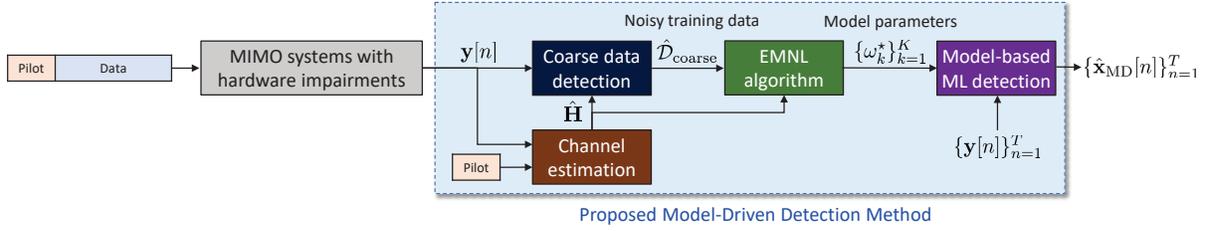, width = 16cm}}\vspace{-3mm}
      \caption{Block diagram of the proposed model-driven detection method in a MIMO system with hardware impairments.} \vspace{-3mm}
      \label{fig:DataDriven}
    \end{figure}

    In this section, we propose a model-driven detection method for MIMO systems with hardware impairments. 
    In this method, to circumvent the limitations of the additive distortion model in \cite{HI : PA 2}, we introduce a generalized Gaussian distortion model that takes into account both deterministic and stochastic distortion caused by hardware impairments. 
    We then present a practical strategy for generating training data to learn this model, which utilizes coarse data detection outputs as noisy training data. 
    To facilitate accurate learning of the generalized model from the noisy training data, we devise an EM algorithm that estimates not only the parameters of the generalized model, but also transition probabilities from noisy labels to true labels. 
    We finally present the ML criterion to detect transmitted data symbols based on the learned model.
    The block diagram of the proposed model-driven detection method is depicted in Fig.~\ref{fig:DataDriven}.


    \subsection{Generalized Distortion Model}\label{Sec:DistortionModel}
    To capture both the deterministic and random effects of non-ideal RF components, we model the distorted received signal at time slot $n$ as 
    \begin{align}
        {\bf y}[n] = {\bf H}{\bf x}[n] + {\bm \mu}({\bf x}[n];{\bf H})+  {\bf d}[n], \label{eq:proposed_model}
    \end{align}    
    where ${\bm \mu}({\bf x}[n];{\bf H}) \in \mathbb{C}^{N_t}$ is the mean of the distortion caused by the non-ideal RF components, and ${\bf d}[n] \sim \mathcal{CN}\big({\bf 0}_{N_r}, {\bf C}({\bf x}[n];{\bf H})\big)$ is the effective noise, and ${\bf C}({\bf x}[n];{\bf H})$ is the covariance of ${\bf d}[n]$ that depends on both ${\bf x}[n]$ and ${\bf H}$. 
    As can be seen in \eqref{eq:proposed_model}, the above model not only considers the presence of a nonzero mean due to the deterministic effects of the non-idea RF components, but also captures the dependency of the effective noise distribution on both the transmitted signal and the channel.   
    We refer to this model as a {\em generalized} distortion model because it generalizes the additive distortion model in \eqref{eq:additive_model}.

    Based on the generalized distortion model, we characterize the conditional distribution of the distorted received signal given a transmitted symbol vector.
    If the MIMO channel does not change during a channel coherence time, both ${\bm \mu}({\bf x}[n];{\bf H})$ and ${\bf C}({\bf x}[n];{\bf H})$ depend solely on the transmitted symbol vector ${\bf x}[n]$ which belongs to the symbol vector set $\mathcal{X}^{N_t} = \{{\bf x}_1, \ldots, {\bf x}_{K}\}$ with $K=|\mathcal{X}^{N_t}|$.
    Motivated by this fact, we define the conditional mean and covariance of ${\bf y}[n]$ given ${\bf x}[n]={\bf x}_k$ as 
    \begin{align}
        {\bm \mu}_k &\triangleq  {\bf H}{\bf x}_k +  {\bm \mu}({\bf x}_k;{\bf H}), \nonumber \\
        {\bf C}_k &\triangleq {\bf C}({\bf x}_k;{\bf H}),
    \end{align} 
    respectively, within each channel coherence time. 
    Then the conditional distribution of ${\bf y}[n]$ given ${\bf x}[n]={\bf x}_k$ is computed as
    \begin{align}\label{eq:Exact_model}
        p\big({\bf y}[n] \big| {\bf x}[n]={\bf x}_k \big) 
        =  \frac{1}{\pi^{N_r}|{\bf C}_k|} 
        \exp\left(-({\bf y}[n]- {\bm \mu}_k)^{\sf H} {\bf C}_k^{-1} ({\bf y}- {\bm \mu}_k)\right).
    \end{align}    
    The above distribution can be simplified if the covariance of the effective noise for ${\bf x}_k$ is approximated by ${\bf C}_k \approx {\nu}_k {\bf I}_{N_r}$ for $\nu_k>0$.
    This approximation is tight when a signal-to-noise ratio (SNR) is low, as can be seen in \eqref{eq:additive_noise}.
    Under this approximation, the conditional distribution in \eqref{eq:Exact_model} is rewritten as
    \begin{align}\label{eq:Gauss_model}
        p\big({\bf y}[n] \big| {\bf x}[n]={\bf x}_k \big) 
        =  \frac{1}{(\pi\nu_k)^{N_r}}\exp\left(-\frac{\Vert {\bf y}[n]-{\bm \mu}_k\Vert^2}{\nu_k}\right) \triangleq p({\bf y}[n]; \omega_k),
    \end{align}    
    where $\omega_k = ({\bm \mu}_k, \nu_k)$. 
    The above characterization implies that the information of the model parameters $\{\omega_k\}_{k=1}^K$ is required at the receiver to perform the optimal ML detection in MIMO systems under hardware impairments.

    \subsection{Training Data Generation Strategy}\label{Sec:Data}
    To train the parameters $\{\omega_k\}_{k=1}^K$ of the generalized model in \eqref{eq:Gauss_model}, the receiver requires a set of training data samples, each of which describes the behavior of the distorted received signal for each symbol vector.
    A conventional strategy to acquire such training data is to send the repetitions of all the symbol vectors as pilot signals \cite{Learning : 1 (SLD), Learning : 2 (SLD)}.
    Unfortunately, this strategy incurs significant training overhead that exponentially increases with the number of transmit antennas and modulation order. 
    To ensure a practical level of training overhead, we develop a training data generation strategy based on conventional channel estimation and data detection, as illustrated in Fig.~\ref{fig:DataDriven}.
    Our strategy works as follows:
    \begin{itemize}
        \item {\bf Step 1 (Pilot-assisted channel estimation):} As in conventional pilot-assisted channel estimation, the transmitter sends pilot signals for channel estimation, while the receiver estimates a channel matrix by utilizing the knowledge of the pilot signals. We denote the estimate channel matrix by $\hat{\bf H}\in\mathbb{C}^{N_r\times N_t}$.
        To be more specific, suppose that $T_{\rm p}$ pilot signals, namely $\{{\bf x}_{\rm p}[n]\}_{n={-T_{\rm p}+1}}^0$, are transmitted from time slot $-T_{\rm p}+1$ to time slot $0$. 
        If we adopt the least-squares (LS) channel estimation method, the estimated channel matrix is given by 
        \begin{align}
            \hat{\bf H} = {\bf Y}_{\rm p} {\bf X}_{\rm p}^{\sf H} \big({\bf X}_{\rm p}{\bf X}_{\rm p}^{\sf H}\big)^{-1}, \label{eq:LS_CE}
        \end{align}
        where ${\bf X}_{\rm p} = \big[{\bf x}_{\rm p}[1], \cdots, {\bf x}_{\rm p}[T_{\rm p}]\big]$ and ${\bf Y}_{\rm p} = \big[{\bf y}[-T_{\rm p}+1], \cdots, {\bf y}[0]\big]$.
        
        \item {\bf Step 2 (Coarse data detection):} Based on the estimate channel, the receiver performs a coarse data detection during data transmission. 
        In this detection, the receiver ignores the effects of hardware impairments as well as channel estimation errors by assuming that ${\bf y}[n] = \hat{\bf H}{\bf x}[n] + {\bf z}[n]$. 
        We denote the index of the coarsely detected symbol vector at time slot $n$ by $\hat{k}[n] \in\{1,\ldots,K\}$.
        For example, if we adopt the ML criterion under the assumption of ${\bf y}[n] = \hat{\bf H}{\bf x}[n] + {\bf z}[n]$, the coarsely detected symbol index $\hat{k}[n]$ is determined as
        \begin{align}
            \hat{k}[n] 
            &=  \underset{k \in \{1,\ldots,K\}}{\argmin}~\Vert {\bf y}[n]-\hat{\bf H} {\bf x}_k\Vert^2.\label{eq:ML_coarse}
        \end{align}

        \item {\bf Step 3 (Training data acquisition):} The receiver assigns a label of $\hat{k}[n]$ to the received signal ${\bf y}[n]$ at time slot $n$.
        Then a training data set is obtained as
         \begin{align}\label{eq:training_data}
            \hat{\mathcal{D}}_{\rm coarse} = \big\{ (\hat{k}[n], {\bf y}[n]) ~|~ n\in\{1,\ldots,T\}\big\}.
        \end{align}
    \end{itemize}
    The most prominent advantage of our strategy is that it requires no extra training overhead beyond the traditional pilot overhead required for MIMO channel estimation. 
    When employing this strategy, however, some training data samples are associated with incorrect labels (i.e., $\hat{\bf x}_{\rm c}[n] \neq {\bf x}[n]$) due to both channel estimation and data detection errors in the presence of hardware impairments.
    Therefore, the receiver needs to learn the parameters $\{\omega_k\}_{k=1}^K$ from these {\em noisy} training data without knowing which data samples have incorrect labels. 

    \subsection{EM Algorithm with Noisy Labels}
    To facilitate accurate learning of the parameters $\{\omega_k\}_{k=1}^K$ from {\em noisy} training data, we develop an EM algorithm, referred to as an {\em EM with noisy labels (EMNL)} algorithm, inspired by a robust neural-network training algorithm in \cite{NLNN}.
    A key assumption of the EMNL algorithm is that a label $\hat{k}$ in noisy training data has its own probability of being associated with a true label $k$.
    This assumption can be mathematically characterized by a {\em label-correcting} channel with a transition matrix ${\bm \Theta}\in\mathbb{R}^{K\times K}$ whose $(i,j)$-th entry represents the conditional probability of the event that a true label is $i$ when observing a noisy label $j$, i.e., 
    \begin{align}
        {\bm \Theta} (i,j) = \mathbb{P}\big(k=i\big|\hat{k}=j\big),~\forall i,j \in\{1,\ldots,K\}.
    \end{align} 
    The EMNL algorithm treats the entries of ${\bm \Theta}$ as unknown parameters and learns these entries as part of parameter learning.
    Note that in the perspective of data detection, ${\bm \Theta} (i,j)$ corresponds to the probability of the event that a transmitted symbol vector is ${\bf x}_k$ when a detected symbol vector is ${\bf x}_{\hat{k}}$. 
    Under the above assumption, the conditional distribution of ${\bf y}[n]$ given $\hat{k}[n]=i$ is computed as
    \begin{align}\label{eq:LF_value}
        p\big(\bold{y}[n]\big|\hat{{k}}[n]=i;\bar{\omega},{\bm \Theta}\big)  
        &= \sum^{K}_{j=1}\mathbb{P}\big(k[n]=j\big|\hat{k}[n]=i;{\bm \Theta}\big)  p\big(\bold{y}[n]  \big|k[n]=j;\bar{\omega}\big) \nonumber \\
        &= \sum^{K}_{j=1} {\bm \Theta} (j,i) p\big(\bold{y}[n] ; \omega_j \big),
    \end{align}
    where $\bar{\omega}= \{({\bm \mu}_k,\nu_k)\}_{k=1}^K$ is a set of the model parameters.

    Based on the above assumption, the EMNL algorithm employs the EM principle to iteratively find the ML estimates of model parameters as well as a transition matrix from noisy training data $\hat{\mathcal{D}}_{\rm coarse}$, while treating true labels as hidden variables.
    Following the EM principle in \cite{NLNN}, the EMNL algorithm aims at maximizing a log-likelihood function of $\bar{\omega}$ and ${\bm \Theta}$ for given $\hat{\mathcal{D}}_{\rm coarse}$, defined as 
    \begin{align}\label{eq:LL_function}
         L(\bar{\omega},{\bm \Theta})
        &=  \sum^T_{n=1} \ln p\big(\bold{y}[n]\big|\hat{{k}}[n];\bar{\omega},{\bm \Theta}\big).
    \end{align}
    To achieve this goal, for every iteration $t \in\{1,\ldots,I_{\rm EMNL}\}$, the EMNL algorithm computes the parameters $\bar{\omega}$ and ${\bm \Theta}$ that maximize the EM-auxiliary function defined as
    \begin{align}\label{eq:EM_aux}
        Q\big(\bar{\omega},{\bm \Theta}; \bar{\omega}^{(t)} ,{\bm \Theta}^{(t)} \big) 
        &= \sum^T_{n=1}\sum^K_{i=1}p\big(k[n]=i \big|{\bf y}[n],\hat{k}[n];\bar{\omega}^{(t)} ,{\bm \Theta}^{(t)} \big)  \nonumber \\
        &~~~\times \big\{\ln p\big({\bf y}[n] \big|k[n]=i;\bar{\omega} \big)
        +\ln p\big(k[n]=i\big|\hat{k}[n]; {\bm \Theta} \big)\big\},
    \end{align}
    where $\bar{\omega}^{(t)}$ and ${\bm \Theta}^{(t)}$ are the model parameters and the transition matrix estimated before iteration $t$. 
    The computation of the parameters is performed by a two-step approach: {\em E-step} and {\em M-step}, as elaborated below.
    
    {\bf E-step:} In the E-step at iteration $t$, the distribution of a true label $k[n]$ is computed based on a noisy training data $(\hat{k}[n],{\bf y}[n])$ and the current parameters as follows: 
    \begin{align}\label{eq:true_label}
        \alpha_{n,i}^{(t)} &\triangleq \mathbb{P}\big(k[n]=i\big|\bold{y}[n],\hat{k}[n];\bar{\omega}^{(t)},{\bm \Theta}^{(t)}\big)  \nonumber  \\
         &= 
         \frac{ \mathbb{P}\big(k[n]=i, \bold{y}[n] \big| \hat{k}[n];\bar{\omega}^{(t)},{\bm \Theta}^{(t)}\big) }
         { p\big(\bold{y}[n] \big| \hat{k}[n];\bar{\omega}^{(t)},{\bm \Theta}^{(t)}\big)} 
         \nonumber  \\
        &\overset{(a)}{=} 
        \frac{ {\bm \Theta}^{(t)}(i,\hat{k}[n])p(\bold{y}[n]; {\omega}_i^{(t)})}
        {\sum_{j=1}^K {\bm \Theta}^{(t)}(j,\hat{k}[n])p(\bold{y}[n]; {\omega}_j^{(t)})}, 
    \end{align}
    where the equality (a) follows from \eqref{eq:LF_value}, and ${\omega}_j^{(t)} = ({\bm \mu}_j^{(t)}, \nu_j^{(t)})$ is the model parameter for label $j$ estimated before iteration $t$.

    {\bf M-step:} In the M-step at iteration $t$, both the model parameters and the transition matrix are updated based on the distribution of the true labels.
    From \eqref{eq:EM_aux}, the model parameters are updated by solving the following problem:
    \begin{align}
        &\underset{\bar{\omega}} \max ~ \sum^T_{n=1}\sum^K_{i=1}  \alpha_{n,i}^{(t)}   \ln p\big({\bf y}[n] \big|k[n]=i;\bar{\omega} \big)  \nonumber \\
        &~\text{s.t.} ~~ \nu_i > 0, \forall i.
    \end{align}
    A closed-form solution of the above problem is readily obtained as
    \begin{align}
        \nu_i^{(t+1)} &= \frac{1}{ N_r\bar{\alpha}_{i}^{(t)}} \sum^T_{n=1}\alpha_{n,i}^{(t)}\Vert\bold{y}[n]-{\bm \mu}_i\Vert^2, \label{eq:nu_update} \\
        \bar{\bm \mu}_i^{(t+1)} &= \frac{1}{\bar{\alpha}_{i}^{(t)} } \sum^T_{n=1} \alpha_{n,i}^{(t)} \bold{y}[n], \label{eq:mu_update}
    \end{align}
    for all $i\in\{1,\ldots,K\}$, where $\bar{\alpha}_{i}^{(t)}= \sum^T_{n=1} \alpha_{n,i}^{(t)}$.

    Similarly, from \eqref{eq:EM_aux}, the transition matrix is updated by solving the following problem:
    \begin{align}
        &\underset{{\bm \Theta}} \max ~ \sum^T_{n=1}\sum^K_{i=1}  \alpha_{n,i}^{(t)}   \ln \mathbb{P}\big(k[n]=i\big|\hat{k}[n]; {\bm \Theta} \big)  \nonumber \\
        &~\text{s.t.} ~~ \sum^K_{i=1}{\bm \Theta}(i,j)=1,\forall j.
    \end{align}
    By simple differentiation with the Lagrange multiplier, a closed-form solution of the above problem is obtained as
    \begin{align}\label{eq:Theta_update}
        {\bm \Theta}^{(t+1)}(i,j) 
        = {\frac{\sum_{n=1}^{T}\alpha_{n,i}^{(t)}\mathbb{I}[\hat{k}[n]=j]} {\sum_{n=1}^{T}\mathbb{I}[\hat{k}[n]=j]}},
    \end{align}
    where $\mathbb{I}\{A\}$ is an indicator function that equals one if $A$ is true and zero otherwise.
    
    {\bf Initialization:} To initialize the EMNL algorithm, we set $\omega_k^{(1)} = ({\bm \mu}_k^{(1)}, \nu_k^{(1)} ) = ( \hat{\bf H}{\bf x}_k, \sigma^2)$ by making an ideal assumption that ${\bf x}_k$ is received via a MIMO channel $\hat{\bf H}$ with no hardware impairments. 
    We also initialize the distribution of a true label from \eqref{eq:initial_true_label} by assuming that ${\bm \Theta}={\bf I}_K$ as follows: 
    \begin{align}\label{eq:initial_true_label}
        \alpha_{n,i}^{(1)} &=
        \frac{ p(\bold{y}[n]; {\omega}_i^{(1)})}
        {\sum_{j=1}^K p(\bold{y}[n]; {\omega}_j^{(1)})}.
    \end{align}
    By applying $\{\alpha_{n,i}^{(1)}\}_{n=1}^T$ to \eqref{eq:Theta_update}, we initialize the transition matrix as
    \begin{align}\label{eq:initial_transition}
        {\bm \Theta}^{(1)}(i,j) = \frac{\sum_{n=1}^T \alpha_{n,i}^{(1)} \mathbb{I}[\hat{k}[n]=j ]}{\sum_{n=1}^{T}\mathbb{I}[\hat{k}[n]=j]}.
    \end{align}

    \subsection{ML Criterion based on the Learned Model}
    After learning the model parameters using the EMNL algorithm, we determine a fine detection rule based on the learned model. 
    By adopting the ML criterion, we estimate the transmitted symbol vector at time slot $n$ as $\hat{\bf x}[n] = {\bf x}_{k^\star[n]}$, where 
    \begin{align}
        k_{\rm MD}^\star[n] 
        &= \underset{k\in \{1,\ldots,K\}}{\argmax}~ p({\bf y}[n]; \omega_k^\star ) \nonumber \\
        &=  \underset{k \in \{1,\ldots,K\}}{\argmax}~\frac{1}{(\pi\nu_k^\star)^{N_r}}\exp\left(-\frac{\Vert {\bf y}[n]-{\bm \mu}_k^\star\Vert^2}{\nu_k^\star}\right), \label{eq:ML_model}
    \end{align}      
    where $\omega_k^\star = ({\bm \mu}_k^\star,\nu_k^\star)$ is the model parameter learned for label $k$. 
    
    The proposed model-driven detection method is summarized in {\bf Procedure~1}, and the block diagram of this method is illustrated in Fig.~\ref{fig:DataDriven}.
    To initialize the EMNL algorithm, we set $\omega_k^{(1)} = ({\bm \mu}_k^{(1)}, \nu_k^{(1)} ) = ( \hat{\bf H}{\bf x}_k, \sigma^2)$ by making an ideal assumption that ${\bf x}_k$ is received via a MIMO channel $\hat{\bf H}$ with no hardware impairments. 
    To improve the numerical stability of the EMNL algorithm, we set ${\alpha}_{n,i}^{(t)} = \tilde{\alpha}_{n,i}^{(t)} / \Sigma_j \tilde{\alpha}_{n,j}^{(t)}$ with $\tilde{\alpha}_{n,i}^{(t)} = \max\{\alpha_{n,i}^{(t)},\epsilon \}$ by introducing an extremely small value $\epsilon \ll 1$ (see Step 5 and Step 10).
    
    \begin{algorithm}[t]
    \caption{The Proposed Model-Driven Detection Method}\label{alg:ModelDriven}
    {\small
        {\begin{algorithmic}[1]
            \REQUIRE Received signals, pilot signals, $\sigma^2$, $I_{\rm EMNL}$, $\epsilon$.
            \ENSURE Detected symbol vectors $\hat{\bf x}[1],\ldots, \hat{\bf x}[T]$.
            \STATE Compute an estimated MIMO channel $\hat{\bf H}$ from channel estimation with pilot signals.
            \STATE Determine a noisy training data set $\hat{\mathcal{D}}_{\rm coarse}$ from coarse data detection. 
            \STATE Set ${\bm \mu}_k^{(1)} = \hat{\bf H}{\bf x}_k$ and $\nu_k^{(1)} = \sigma^2$, $\forall k$. 
            \STATE Initialize $\alpha_{n,i}^{(1)}$ from \eqref{eq:initial_true_label}, $\forall n,i$.
            \STATE Set ${\alpha}_{n,i}^{(1)} = \tilde{\alpha}_{n,i}^{(1)} / \Sigma_j \tilde{\alpha}_{n,j}^{(1)}$, where $\tilde{\alpha}_{n,i}^{(1)} = \max\{\alpha_{n,i}^{(1)},\epsilon \}$, $\forall n,i$.
            \STATE Initialize ${\bm \Theta}^{(1)}(i,j)$ from \eqref{eq:initial_transition}, $\forall i,j$.
            \FOR {$t=1$ to $I_{\rm EMNL}$}
            \STATE \!\!\!{\bf E-step:}
            \STATE Compute $\alpha_{n,i}^{(t)}$ from \eqref{eq:true_label}, $\forall n,i$.
            \STATE Set ${\alpha}_{n,i}^{(t)} = \tilde{\alpha}_{n,i}^{(t)} / \Sigma_j \tilde{\alpha}_{n,j}^{(t)}$, where $\tilde{\alpha}_{n,i}^{(t)} = \max\{\alpha_{n,i}^{(t)},\epsilon \}$, $\forall n,i$.
            \STATE \!\!\!{\bf M-step:} 
            \STATE Update $\nu_i^{(t+1)}$ and ${\bm \mu}_i^{(t+1)}$ from \eqref{eq:nu_update} and \eqref{eq:mu_update}, $\forall i$.
            \STATE Update ${\bm \Theta}^{(t+1)}(i,j)$ from \eqref{eq:Theta_update}, $\forall i,j$.
            \ENDFOR
            \STATE Determine $k_{\rm MD}^\star[n] 
            = \underset{k}{\argmax}~ p({\bf y}[n]; \omega_k^\star)$, $\forall n$.
            \STATE Set $\hat{\bf x}_{\rm MD}[n] = {\bf x}_{k_{\rm MD}^\star[n]}$, $\forall n$.
        \end{algorithmic}}}
    \end{algorithm}

\section{Proposed Data-Driven MIMO Detection}\label{Sec:DataDriven}   
    Although the proposed model-driven detection method in Sec.~\ref{Sec:ModelDriven} adopts a generalized Gaussian modeling approach to compensate for hardware impairments, a major limitation of this approach is its inability to capture the effects of non-Gaussian distortion. Unfortunately, the impacts of some hardware impairments, such as nonlinear saturation in PAs and quantization errors in low-resolution ADCs, do not follow Gaussian distributions. As a result, our model-driven method may suffer from a model mismatch in the presence of non-Gaussian distortion caused by these hardware impairments.
    To address the aforementioned limitation of the model-driven detection method, in this section, we propose a data-driven detection method based on DL, which does not rely on any assumptions regarding the distribution of the distorted received signals.

    \subsection{Fundamental Idea: Deep Learning with Noisy Labels}\label{Sec:Noisy_Labels}
        The maximum a-posteriori probability (MAP) criterion for the MIMO detection problem is given by
        \begin{align}
            k_{\rm MAP}^{\star}[n] 
            &= \underset{i\in \{1,\ldots,K\}}{\argmax}~ \mathbb{P}({\bf x}[n] = {\bf x}_i|{\bf y}[n]) \nonumber \\
            &= \underset{i\in \{1,\ldots,K\}}{\argmax}~ \mathbb{P}(k[n]=i|{\bf y}[n]) , 
        \end{align}
        where $\mathbb{P}(k[n]=i|{\bf y}[n])$ is the APP of having ${\bf x}[n] = {\bf x}_i$ when observing ${\bf y}[n]$. 
        The above MAP criterion is well-known for its optimality in terms of minimizing the decision error probability.
        From the Bayes' rule, the APP $\mathbb{P}(k[n]=i|{\bf y}[n])$ is rewritten as
        \begin{align}\label{eq:APP_def}
            \mathbb{P}(k[n]=i|{\bf y}[n])  
            &= \frac{p({\bf y}[n] | k[n]=i)\mathbb{P}(k[n]=i)}{p({\bf y}[n]) } \nonumber \\
            &= \frac{p({\bf y}[n] | k[n]=i)\mathbb{P}(k[n]=i)}{\sum_{j=1}^K  p({\bf y}[n] | k[n]=j)\mathbb{P}(k[n]=j)}.
        \end{align}
        Then, an APP vector for the received signal ${\bf y}[n]$ can be defined as 
        \begin{align}\label{eq:APP_vec}
            {\bf p}[n] =  
            \begin{bmatrix} 
            \mathbb{P}(k[n]=1|{\bf y}[n]) \\ \vdots \\ \mathbb{P}(k[n]=K|{\bf y}[n]) \end{bmatrix} \in [0,1]^{K}.
        \end{align}
        As can be seen from \eqref{eq:APP_def} and \eqref{eq:APP_vec}, an exact characterization of the APP vector ${\bf p}[n]$ requires perfect knowledge of $p({\bf y}[n] | k[n]=i)$ for all $i\in\{1,\ldots,K\}$. 
        However, as we have already discussed, the conditional distribution $p({\bf y}[n] | k[n]=i)$ is unknown in the presence of hardware impairments; thereby, the exact characterization of ${\bf p}[n]$ is infeasible for MIMO systems under hardware impairments.

\begin{figure}[t]
          \centering \vspace{-3mm}
               {\epsfig{file=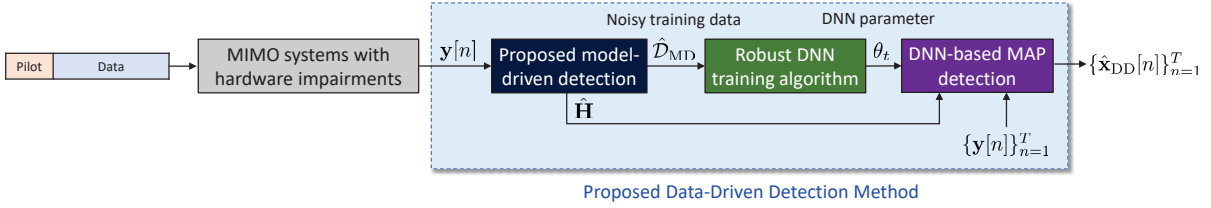, width = 16cm}}\vspace{-3mm}
          \caption{Block diagram of the proposed data-driven detection method in a MIMO system with hardware impairments.} \vspace{-3mm}
          \label{fig:Block2}
        \end{figure}

        To circumvent the aforementioned challenge, in the proposed data-driven method, we employ a DNN as a nonlinear function approximator to approximate the APP vector in \eqref{eq:APP_vec} for each distorted received signal. 
        In particular, we aim at determining the DNN with parameter $\theta$ that satisfies 
        \begin{align}\label{eq:DNN_approx}
            {\bf p}[n] \approx f_{\theta}^{\rm DNN}({\bf y}[n],\hat{\bf H}),~~\forall n\in\{1,\ldots,T\}.
        \end{align}
        Note that the input of our DNN is not only the received signal ${\bf y}[n]$, but also the estimated channel matrix $\hat{\bf H}$ that can be obtained by employing conventional pilot-assisted channel estimation. 
        Since the conditional distribution $p({\bf y}[n] | k[n]=i)$ depends on the channel matrix ${\bf H}$, utilizing $\hat{\bf H}$ as an additional input may provide some useful information to approximate the APP vector.
        Training the DNN parameter $\theta$ in \eqref{eq:DNN_approx} requires training samples that can represent the relation between the distorted received signal ${\bf y}[n]$ and the transmitted symbol index $k[n]$.
        These training samples can be readily attained using the training data acquisition strategy in Sec.~\ref{Sec:Data}, but in this case, the resulting training samples can be highly unreliable.
        To resolve this problem, we first refine the labels of these unreliable training data samples by applying the model-driven detection method in Sec.~\ref{Sec:ModelDriven}.
        We then generate a training data set as $\hat{\mathcal{D}}_{\rm MD}=\{({\bf e}_{\rm MD}[n],{\bf y}[n])\}^T_{n=1}$, where ${\bf e}_{\rm MD}[n]$ is a one-hot-encoded target vector which consists of all zeros except for the ${k}_{\rm MD}^{\star}[n]$-th component which has the value of $1$, and ${k}_{\rm MD}^{\star}[n]$ is the estimated symbol index for ${\bf y}[n]$ determined by the model-driven method.
        The block diagram of the proposed data-driven detection method with the above strategy is depicted in Fig.~\ref{fig:Block2}.

        A typical approach for training the DNN parameter $\theta$ is to employ a mini-batch gradient descent algorithm in which the DNN parameter is iteratively updated in the direction of the negative gradient of the empirical loss of mini-batch samples.
        Let $\mathcal{B} \subset \{1,\ldots,T\}$ be the index set of the samples in the mini-batch drawn from $\hat{\mathcal{D}}_{\rm MD}$.
        Then, the original update equation of the mini-batch gradient descent algorithm for the DNN parameter $\theta_{t}$ is given by
        \begin{align}\label{eq:Update1}
            \theta_{t+1} = \theta_{t} - \eta \nabla \left(\frac{1}{\vert \mathcal{B}\vert}\sum_{n \in \mathcal{B}} \ell({\bf e}_{\rm MD}[n],{\bf y}[n];\theta_t)\right),
        \end{align}
        where $\eta > 0$ is a learning rate, $\nabla (\cdot)$ is a gradient operator, and $\ell({\bf e},{\bf y};\theta_t)$ is a loss of the sample $({\bf e},{\bf y})$ computed for the DNN parameter $\theta_t$.
        Unfortunately, in the proposed data-driven method, the training samples in $\hat{\mathcal{D}}_{\rm MD}$ are still {\em noisy} not only due to the noise in the communication system, but also due to the model mismatch under non-Gaussian distortion.
        If we simply train the DNN parameter from \eqref{eq:Update1} using the noisy training samples, the DNN can easily fit the noisy labels because it has the ability to memorize the relationships of the training samples \cite{Deep : Memorization1}.
        Furthermore, even well-trained DNNs can be susceptible to small amounts of added noise on the training data \cite{Deep : Fool}.
        These effects cause the DNN to be biased towards the noisy training data, resulting in overfitting that significantly degrades the generalization performance of the DNN \cite{Deep : Generalization}.

        \subsection{Robust Training Algorithm}\label{Sec:Robust_Algorithm}
        To address the overfitting problem caused by the noisy training samples, we devise a robust DNN training algorithm by integrating the techniques suggested in \cite{Deep : SELFIE} and \cite{Deep : Self-adaptive}.
        Our algorithm involves a warm-up period, sample selection, and loss correction.
        Details of each strategy are described below.

        {\bf Warm-Up Period:} 
        The parameters of the DNN are initialized with arbitrary values that are far from the final desired values. Therefore, during the initial training phase, the parameters of the DNN are unstable and may change rapidly.
        These unstable changes tend to decrease as the training progresses due to the memorization effect \cite{Deep : Memorization1}, but cause continuous changes in the DNN output before stabilizing.
        A continuously changing DNN output changes the value of the loss significantly, which also changes the direction of the gradient.
        Therefore, during the initial training phase, the DNN may not be able to determine the ideal direction of parameter updates, making it challenging to achieve good performance even when employing some robust training strategies.
        To address this problem, we define the first $I_{\rm warmup}$ epochs of the training process as the warm-up period. During this period, we train the DNN parameter using all the noisy training samples in a default manner, so that the randomly initialized parameters of the DNN are stabilized and placed near the desired values.
        Although we use all the noisy training samples during the warm-up period, it is reported that the DNN tends to first learn patterns from easy samples (i.e., low-loss samples) in the initial learning phase and then learns details from difficult samples (i.e., high-loss samples) as training progresses \cite{Deep : Memorization1}.
        Since the warm-up period is part of the initial training phase, even if the training samples are noisy and include false labels, the DNN may focus on learning patterns from clean and easy samples without being significantly affected by false and difficult samples.
        In this context, the warm-up period allows the DNN to be stabilized and prepared before applying robust training strategies. 
        

        \begin{figure}[t]
            \centering 
    		\subfigure[$\text{Iteration}=0$]
    		{\epsfig{file=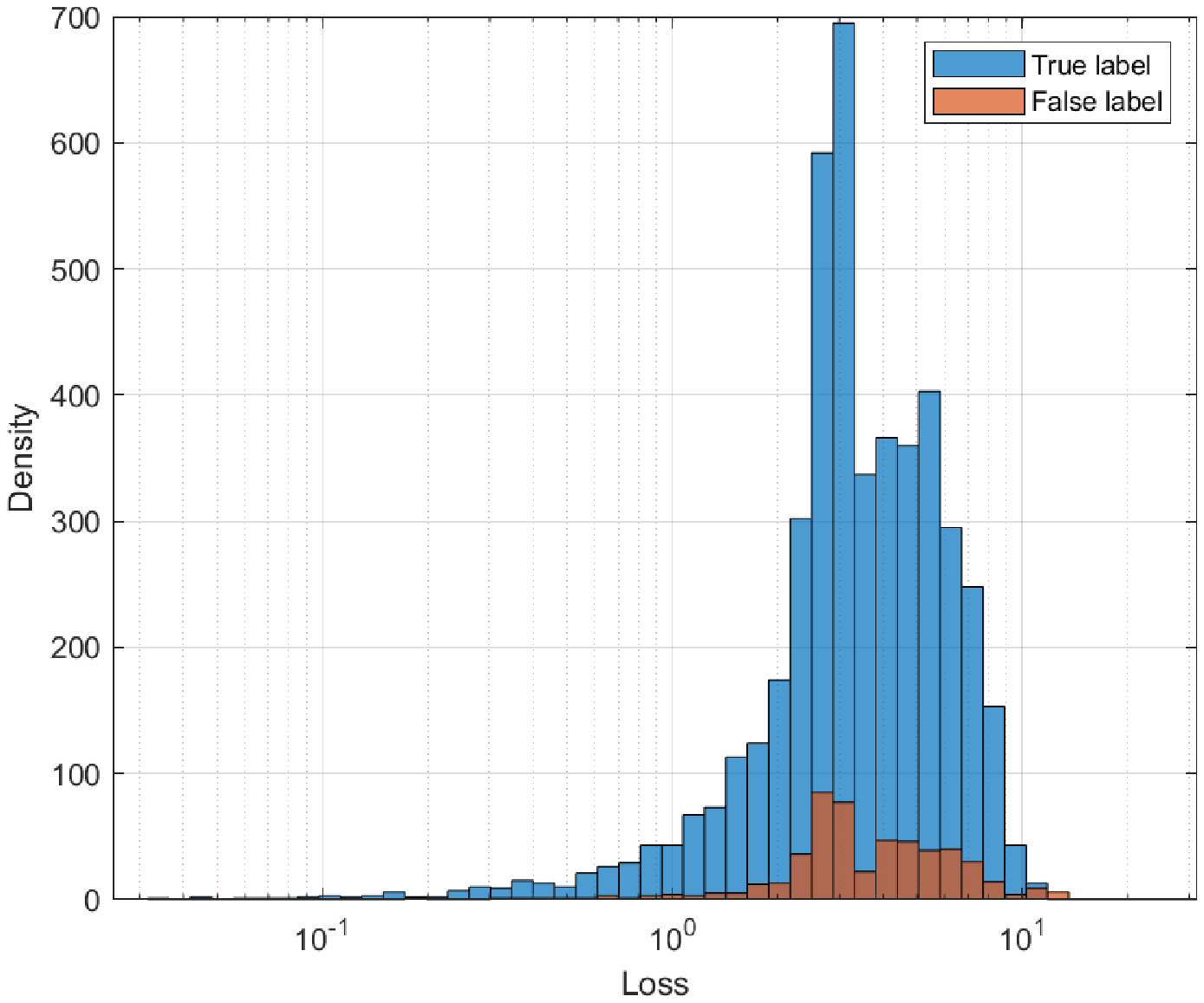, width=8cm}}
            \centering 
                \subfigure[$\text{Iteration}=40$]
                {\epsfig{file=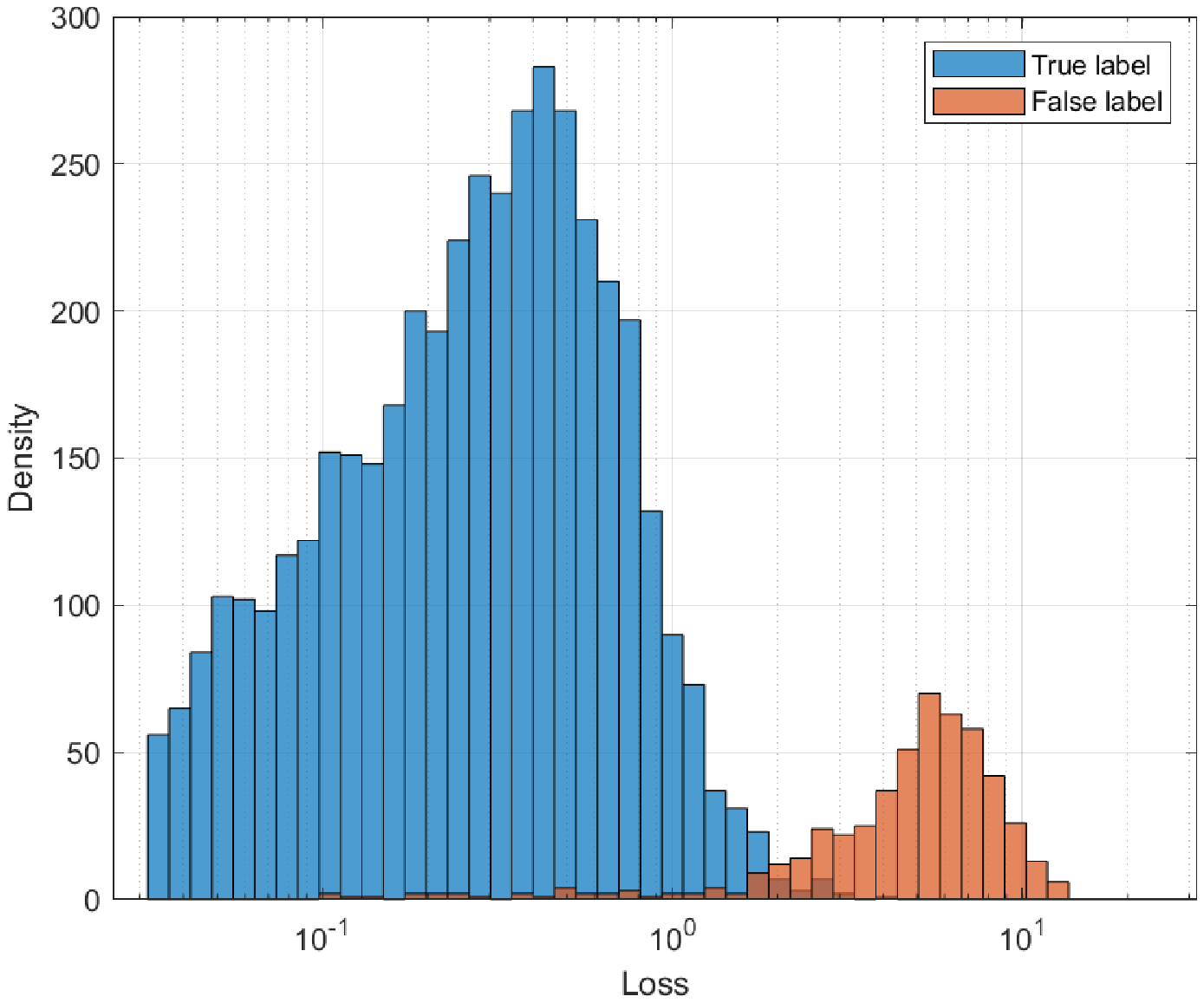, width=8cm}}
            \caption{Histogram comparison of log-scale loss distributions between true and false training samples when noise rate is $10\%$ before and after the warm-up period.}  
            \label{fig:Loss_Comparison}
        \end{figure}

        {\bf Sample Selection:} 
        The key idea of the sample selection strategy is to categorize the mini-batch samples into three types: (i) false samples, (ii) clean samples, and (iii) reassignable samples. 
        In noisy training sets, it is generally difficult to accurately distinguish between false and true samples. However, it is still possible to roughly identify {\em some} unreliable samples based on the loss after the warm-up period \cite{Deep : Coteaching, Deep : MentorNet, Deep : Coregularization, Deep : Disagreement}.  
        To give an idea about this strategy, in Fig.~\ref{fig:Loss_Comparison}, we depict the histogram of the log-scale loss distributions of the true and false samples before and after the warm-up period. Fig.~\ref{fig:Loss_Comparison} shows that most of the clean training samples have a lower loss than the noisy training samples after the warm-up period.
        Motivated by these observations, we first compute the losses of the mini-batch samples w.r.t. the DNN parameter $\theta$ obtained after the warm-up period. 
        We then categorize the $|\mathcal{B}|\tau$ samples with the highest losses as {\em false} samples. 
        Let $\mathcal{F} \subset \mathcal{B}$ be a set of the indexes of the false samples selected by the above strategy. 
        Unfortunately, as shown in Fig.~\ref{fig:Loss_Comparison}, losses of some false samples can be lower than those of true samples even after the warm-up period. 
        For this reason, the samples in the index set $\mathcal{B}\setminus \mathcal{F}$ (which excludes the false samples based on the losses) consists mostly of clean samples, but may also include false samples. 
        To address this issue, we further split the samples in the index set $\mathcal{B}\setminus \mathcal{F}$ into two types: {\em clean} and {\em reassignable} samples.
        To this end, we define a label confidence for each sample by utilizing the prediction output of the DNN. 
        Consider the prediction output of the DNN with $\theta_t$ for the input ${\bf y}[n]$, given by 
        \begin{align}\label{eq:DNN_approx}
            \hat{\bf p}[n] = f_{\theta_t}^{\rm DNN}({\bf y}[n],\hat{\bf H}).
        \end{align}
        If the output nodes of the DNN employ the softmax activation functions, which is typical for a classification task, the prediction output $\hat{\bf p}[n]$ for the input ${\bf y}[n]$ can be interpreted as the approximation of the APP vector ${\bf p}[n]$ defined in \eqref{eq:APP_vec}. 
        Then the $k$-th entry of $\hat{\bf p}[n]$, namely $\hat{p}_k[n]$, represents the probability of the event that the input ${\bf y}[n]$ is associated with class $k$.
        This implies that the higher the maximum value of $\hat{\bf p}[n]$, the more confident the DNN is about the prediction for ${\bf y}[n]$.
        A straightforward strategy utilizing this fact is to define the label confidence using the maximum value of $\hat{\bf p}[n]$ (i.e., $\max_k \hat{p}_k[n]$).
        However, the label confidence that solely relies on the current prediction may impose undesired bias towards the current belief of the DNN and therefore lead to the overfitting problem.
        This problem becomes more severe in the early-to-middle stage of the training as the prediction output in this stage is not only unreliable but also unstable in general. 
        To address the above problem, we consider a {\em target} vector defined as an exponential moving average (EMA) of the prediction output, as suggested in \cite{Deep : Self-adaptive}.
        Let ${\bf t}[n] \in [0,1]^K$ be the target vector for ${\bf y}[n]$ such that $\sum_{k}t_k[n] =1$, where $t_k[n]$ is the $n$-th entry of ${\bf t}[n]$. 
        Once the prediction output of the DNN is obtained as $\hat{\bf p}[n]$ for the input ${\bf y}[n]$, the target vector ${\bf t}[n]$ is updated according to the EMA update rule:
        \begin{align}\label{eq:target_update}
            {\bf t}[n] \leftarrow \alpha \hat{\bf p}[n] + (1-\alpha){\bf t}[n],
        \end{align}
        where $\alpha\in [0,1]$ is a smoothing factor that controls the weight on the prediction output of the DNN.
        The update rule in \eqref{eq:target_update} shows that the target vector smoothens the prediction outputs during the training; thereby, it can alleviate the instability of the DNN output in the early-to-middle stage of the training.
        The update rule of the target vector also implies that the maximum value of ${\bf t}[n]$, namely $\max_k t_k[n]$, can be close to one only if the DNN output consistently shows strong confidence for a particular class during the training process. 
        Motivated by these observations, we define the label confidence  $\omega[n]$ as the maximum value of the target vector ${\bf t}[n]$, i.e.,
        \begin{align}
            \omega[n]  = \max_k t_k[n],
        \end{align}
        as done in \cite{Deep : Self-adaptive}.
        As discussed above, we have $\omega[n] \approx 1$ only if the DNN output $\hat{\bf p}[n]$ consistently shows strong confidence for a particular class during the training process.
        Considering this fact, for $n \in \mathcal{B}\setminus \mathcal{F}$, we treat the sample $({\bf t}[n], {\bf y}[n])$ as clean if $\omega[n] > 1-\epsilon$ for an extremely small constant $\epsilon \ll 1$.
        Consequently, we categorize the samples in the mini-batch that are not categorized as false or clean samples as {\em reassignable} samples.
        Let $\mathcal{C}$ and $\mathcal{R}$ be a set of the indexes of the clean and reassignable samples, respectively.
        Then, our selection strategy implies that for $n \in \mathcal{B}\setminus \mathcal{F}$, we have $n\in\mathcal{C}$ if $\omega[n] > 1-\epsilon$ and $n\in\mathcal{R}$ otherwise. 

        {\bf Loss Correction:} 
        In the original parameter update equation in \eqref{eq:Update1}, the losses of all the training samples are treated {\em equally} because they are believed to be true samples. 
        However, if training samples are noisy, it is more natural to treat the loss of each sample differently because the losses of false samples may provide undesirable biases in the parameter update.
        Motivated by this observation, in the loss correction strategy, we put forth a loss function for the noisy training samples, so that the loss of each sample is treated in a different manner according to the category of the sample determined by the sample selection strategy. 
        Our loss function is straightforward for the false and clean samples. 
        We ignore the losses of the false samples, while treating the losses of the clean samples as done in the original parameter update equation.
        Our challenge lies in how to treat the losses of reassignable samples which can be either true or false. 
        The strategy suggested in \cite{Deep : SELFIE} is to replace the noisy label of the reassignable sample with a new label determined by the current model prediction; then, the losses computed with the re-assigned labels are treated as if they are the losses of the clean samples. 
        When employing this strategy, however, the direction of the gradient descent may be undesirable because the reassigned labels are still unreliable and potentially erroneous.
        This problem can be more severe for the reassignable samples with low label confidences. 
        To resolve this problem, as suggested in \cite{Deep : Self-adaptive}, we compute the loss of the reassignable sample using a {\em target} vector in \eqref{eq:target_update}, instead of the reassigned label. 
        In addition, for the loss of the reassignable sample with index $n\in\mathcal{R}$, we assign a weight  based on the label confidence $\omega[n]$, so as to alleviate the undesirable impact of the reassignable samples with low confidences. 
        From our strategies for computing the losses of different types of samples, the parameter update equation in \eqref{eq:Update1} is modified as
        \begin{align}\label{eq:modified_update}
            \theta_{t+1} = \theta_{t} - \eta \nabla \left(\frac{1}{\vert \mathcal{C} \vert + \sum_{n \in \mathcal{R}}  t_{\rm max}[n] }\left(\sum_{n \in \mathcal{C}} \ell ({\bf t}[n],{\bf y}[n];\theta_t) +   \sum_{n \in \mathcal{R}} t_{\rm max}[n]  \ell ({\bf t}[n],{\bf y}[n];\theta_t)  \right)\right).
        \end{align}
        As can be seen in \eqref{eq:modified_update}, our loss function pays less attention to the reassignable samples with low label confidences and learns more from clean samples. 
        Meanwhile, the reassignable samples with high label confidences are treated very similarly to the clean samples.
        It is also shown that the losses of the false samples are completely excluded from the parameter update.

        \begin{algorithm}[t]
            \caption{The Proposed Data-Driven Detection Method}\label{alg:DataDriven}
            {\small
                {\begin{algorithmic}[1]
                    \REQUIRE Received signals, pilot signals, $\sigma^2$, $I_{\rm epochs}$, $N_{\rm batch}$, $\tau$, $\alpha$, $\epsilon$.
                    \ENSURE Detected symbol vectors $\hat{\bf x}_{\rm DD}[1],\ldots, \hat{\bf x}_{\rm DD}[T]$.
                    \STATE Generate a noisy training data set $\hat{\mathcal{D}}_{\rm MD}$ from {\bf Procedure~\ref{alg:ModelDriven}}.  
                    \STATE Initialize the DNN parameter $\theta_1$ and set $t=1$ and  ${\bf t}[n] = {\bf e}_{\rm MD}[n]$, $\forall n$.
                    \FOR {$i = 1$ to $I_{\rm epochs}$} 
                    \STATE Randomly partition $\hat{\mathcal{D}}_{\rm MD}$ into $N_{\rm batch}$ mini-batches. 
                    \FOR {$j = 1$ to $N_{\rm batch}$}
                    \STATE Set $\mathcal{B}$ as the indexes of the samples in the $j$-th mini-batch.
                    \IF {$i\leq I_{\rm warmup}$}  
                    \STATE Update $\theta_{t+1}$ from \eqref{eq:Update1} and $t\leftarrow t+1$.
                    \ELSE
                    \STATE Compute $\ell({\bf t}[n],{\bf y}[n];\theta_t)$, $\forall n \in\mathcal{B}$.
                    \STATE Set $\mathcal{F}$ as a set of the indexes of the $|\mathcal{B}|\tau$ samples with the highest losses.
                    \STATE Set $\mathcal{C}=\emptyset$ and $\mathcal{R}=  \mathcal{B}\setminus \mathcal{F}$.
                    \FOR {\textbf{each} $n \in\mathcal{B}\setminus \mathcal{F}$}
                    \STATE $\hat{\bf p}[n] = f_{\theta_t}^{\rm DNN}({\bf y}[n],\hat{\bf H})$.
                    \STATE ${\bf t}[n] \leftarrow \alpha   \hat{\bf p}[n] + (1-\alpha){\bf t}[n]$.
                    \STATE $\omega[n]  = \max_k t_k[n]$.
                    \IF {$\omega[n] > 1 - \epsilon$}
                    \STATE  Update $\mathcal{C} \leftarrow \mathcal{C}\cup \{n\}$ and $\mathcal{R} \leftarrow \mathcal{R} \setminus \{n\}$.
                    \ENDIF
                    \ENDFOR
                    \STATE Update $\theta_{t+1}$ from \eqref{eq:modified_update} and $t\leftarrow t+1$.
                    \ENDIF                   
                    \ENDFOR
                    \ENDFOR
                    \STATE Determine $k_{\rm DD}^{\star}[n] 
                    = \underset{k}{\argmax}~\hat{p}_k[n]$ from $\hat{\bf p}[n] = f_{\theta_t}({\bf y}[n],\hat{\bf H})$, $\forall n$.
                    \STATE Set $\hat{\bf x}_{\rm DD}[n] = {\bf x}_{k_{\rm DD}^{\star}[n]}$, $\forall n$.
                \end{algorithmic}}}
        \end{algorithm}
                
        \subsection{MAP Criterion based on the Trained DNN}
        After training the DNN parameters using the robust training algorithm, we determine the MAP detection rule based on the trained DNN. 
        By adopting the MAP criterion, we estimate the transmitted symbol vector at time slot $n$ as $\hat{\bf x}_{\rm DD}[n] = {\bf x}_{k_{\rm DD}^{\star}[n]}$, where 
        \begin{align}
            k_{\rm DD}^{\star}[n] 
            &= \underset{k\in \{1,\ldots,K\}}{\argmax}~f_{\theta^{\star}}^{\rm DNN}({\bf y}[n],\hat{\bf H}), \label{eq:ML_free}
        \end{align}      
        where $\theta^{\star}$ is the DNN parameter after the training process.
        The proposed data-driven detection method is summarized in {\bf Procedure~2}, and the block diagram of this method is illustrated in Fig.~\ref{fig:Block2}.

    \section{Simulation Results}\label{Sec:Simul}
    In this section, using simulations, we demonstrate the superiority of the proposed detection methods in MIMO systems with hardware impairments.
    In the simulation, 4-QAM and $T=500$ are assumed, and the SNR of the system is defined as ${\sf SNR} = N_{t}/\sigma^2$.
    To capture the time-varying nature of the wireless channel, we assume that the MIMO channel matrix at time slot $n$ is given by
    \begin{align}
        {\bf H}[n] = \zeta {\bf H}[n-1] +\sqrt{1-\zeta^2}{\bf G}[n],~\forall n\in\{2,\ldots,T\},
    \end{align}
    where $\zeta \in [0,1]$ is a temporal-correlation coefficient, and ${\bf G}[n]$ is an evolution matrix at time slot $n$ whose entries are independently drawn from $\mathcal{CN}(0,1)$.
    We also assume that the entries of the initial channel ${\bf H}[1]$ are independently drawn from $\mathcal{CN}(0,1)$.
    To model various effects of hardware impairments, we consider two scenarios: 
    \begin{itemize}
        \item {\bf Additive Distortion Scenario:} 
        In this scenario, we consider the additive distortion model in \cite{HI : PA 2} with time-invariant channels (i.e., ${\bf H}[n] = {\bf H},~\forall n$). Under this setting, the baseband received signal at time slot $n$ is given by
	\begin{align}\label{eq:additive_model2}
            {\bf y}[n] = {\bf H}{\bf x}[n] + {\bf z}_{\rm eff}[n],
        \end{align}
        where the distribution of ${\bf z}_{\rm eff}[n]$ is assumed to be
        \begin{align}\label{eq:additive_noise}
            {\bf z}_{\rm eff}[n]  \sim \mathcal{CN}\big({\bf 0}_{N_r},  (\kappa_{\rm tx} + \kappa_{\rm rx}){\bf H}{\bf H}^{\sf H} + {\sigma}^2{\bf I}_{N_{r}} \big).
        \end{align}
        We set $\kappa_{\rm tx} = \kappa_{\rm rx} = 0.05^2$, as done in \cite{HI : PA 2}.
        
        \item {\bf Realistic Distortion Scenario:} 
        In this scenario, we consider the realistic distortion model in \cite{Learning : 3 (ELM)} in which non-ideal PAs are employed at the transmitter and low-resolution ADCs are employed at the receiver.
        The distortion function at the transmitter is expressed as
         $f_{\rm tx}({\bf x}[n]) = {\sf PA}({\bf x}[n])$, where ${\sf PA}(x) =  A(\vert x\vert)e^{j\{\angle(x) + \Phi(\vert x\vert)\}}$, $\angle(x)$ is the angle of $x$, 
        \begin{align}
            A(\vert x\vert) &= \frac{\alpha_a\vert x\vert}{1+\epsilon_a\vert x\vert^2},~~~
            \Phi(\vert x\vert) = \frac{\alpha_\phi \vert x\vert^2}{1+\epsilon_\phi\vert x\vert^2},
        \end{align}
        and $\alpha_a$, $\epsilon_a$, $\alpha_\phi$, $\epsilon_\phi$ are PA-dependent parameters.
        The distortion function at the receiver is expressed as $f_{\rm rx}({\bf r}[n]) = {\sf ADC}({\bf r}[n])$, where $y_k={\sf ADC}(r)$ if $b_{k-1} < r \leq b_k$, $y_k$ is the $k$-th quantization output of the ADC, and $b_k$ is the $k$-th quantization boundary of the ADC.
        We choose $y_k = -1.75+(k-1)/2$ for $k\in\{1,\ldots,2^B\}$ and $b_{k} = (y_k+y_{k+1})/2$ for $k\in\{1,\ldots,2^{B}-1\}$.
        We set $(\alpha_a,\epsilon_a,\alpha_\phi,\epsilon_\phi)=(1.96,0.99,2.53,2.82)$ 
 as done in \cite{PA : 1} and $B=3$. 
        In this scenario, we consider both time-invariant and time-varying channels. 
        When assuming the time-invariant channels, we set $\zeta=1$, and when assuming the time-varying channels, we set $\zeta=0.98$. 
    \end{itemize}

    \begin{figure}[t]
        \centering 
        {\epsfig{file=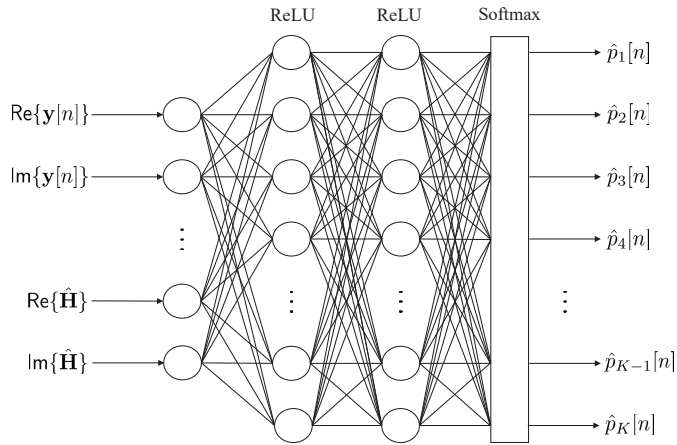, width=9cm}} \vspace{-3mm}
        \caption{Illustration of the DNN architecture adopted in the proposed data-driven detection method.}\vspace{-3mm}
        \label{fig:Neural Network}
    \end{figure}

	For performance comparisons, we consider the following detection methods:
	\begin{itemize}
	    \item {\bf Proposed (Model-Driven):} 
            This is the proposed model-driven detection method summarized in {\bf Procedure~\ref{alg:ModelDriven}}. For the training data generation strategy, we adopt the conventional LS method in \eqref{eq:LS_CE} with $4$ pilot signals (i.e., $T_{\rm p}=4$) for channel estimation and adopt a coarse ML criterion in \eqref{eq:ML_coarse} for coarse data detection. 
            We set $I_{\rm EMNL}=20$ and $\epsilon = 10^{-8}$.

	    \item {\bf Proposed (Data-Driven):} This is the proposed data-driven detection method summarized in {\bf Procedure~\ref{alg:DataDriven}}. For training data generation, we adopt the proposed model-driven detection method specified above. For a DNN architecture, we consider a fully-connected multi-layer perceptron illustrated in Fig.~\ref{fig:Neural Network}. Two hidden layers consist of $100$ nodes with the ReLU activation function and the output layer consists of $K$ nodes with the Softmax activation function.
        The DNN is trained by Adam optimizer with the cross-entropy loss function.
        We set $I_{\rm epochs} = 100$, $I_{\rm warmup}=40$, $N_{\rm batch} = 4$, $\tau = 0.1$, $\alpha = 0.1$, and $\epsilon = 10^{-8}$. 
        We also consider a learning rate scheduler that divides the initial learning rate of $\eta = 0.01$ by $5$ at $50\%$ and $75\%$ of the total number of epochs.

	    \item {\bf Approx. MLD + LS CE:} 
            This is a conventional ML detection method which ignores the distortion caused by hardware impairments. The detection criterion for this method is given in \eqref{eq:ML_coarse}.
            To determine $\hat{\bf H}$ in \eqref{eq:ML_coarse}, channel estimation based on the LS method in \eqref{eq:LS_CE} is adopted with $4$ pilot signals (i.e., $T_{\rm p}=4$).
     
	    \item {\bf DNN + EMNL:}
            This is a slight modification of the DNN-based classifier developed in \cite{NLNN}. 
            This classifier utilizes an EM-based DNN training algorithm to learn both the parameters of the DNN and the transition probabilities of noisy labels, which is similar to the EMNL algorithm. 
            We adopt the same DNN architecture as that used in the proposed data-driven method.  
            We employ the above classifier as a MIMO detection method. 
            

	    \item {\bf Adaptive ELM:} 
            This is an ELM-based detection method developed in \cite{Learning : 3 (ELM)}. We particularly adopt the online-sequential ELM method in \cite{Learning : 3 (ELM)} which initially trains the network using pilot signals and then updates the network for every time slot via online training using a true training data $({\bf x}[n], {\bf y}[n])$ at time slot $n$. 
            
	\end{itemize}

    \begin{figure}[t]
		\centering 
		\epsfig{file=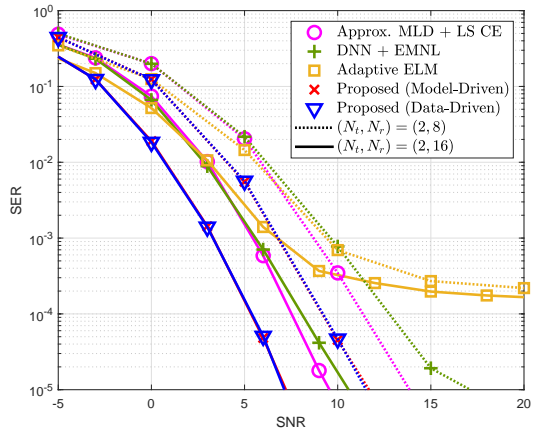, width=8cm}
   \vspace{-3mm}
		\caption{SER comparison of various detection methods for the additive distortion scenario with time-invariant channels when $(N_t,N_r) \in \{(2,8),(2,16)\}$.}  \vspace{-3mm}
		\label{fig:EM_SER}
    \end{figure}
        Fig.~\ref{fig:EM_SER} compares the SERs of various detection methods for the additive distortion scenario with time-invariant channels when $(N_t,N_r) \in \{(2,8),(2,16)\}$.
        Fig.~\ref{fig:EM_SER} shows that the proposed detection methods outperform the existing detection methods for all SNR values.
        It should be noted that the generalized Gaussian model adopted in the proposed model-driven method perfectly matches with the additive distortion scenario in which the distortion caused by hardware impairments is modeled as an additive Gaussian signal. 
        For this reason, Fig.~\ref{fig:EM_SER} shows that the proposed data-driven method does not provide additional performance gain over the model-driven method. 
        Meanwhile, the performance gain of the model-driven method over the existing detection methods implies that the EMNL algorithm effectively finds proper model parameters even from noisy training data. 
        Although DNN + EMNL also adopts a similar EMNL algorithm designed for noisy training data, it shows worse performance than the proposed model-driven method because DNN is more vulnerable to an overfitting problem than our generalized Gaussian model.
        Fig.~\ref{fig:EM_SER} also shows that adaptive ELM has an error floor due to the suboptimality of the ELM approach, even if it uses the true training data during the online training.

    \begin{figure}[t]
		\centering 
		\epsfig{file=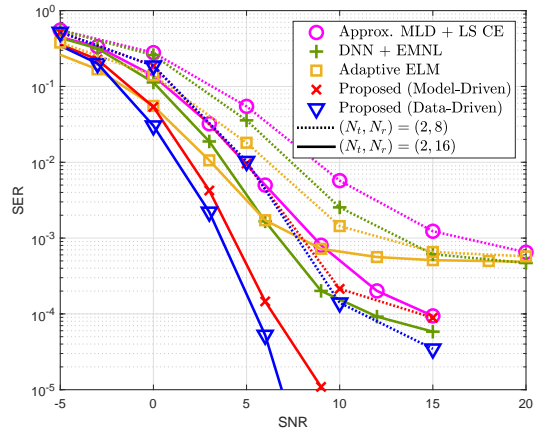, width=8cm}
        \vspace{-3mm}
		\caption{SER comparison of various detection methods for the realistic distortion scenario with time-invariant channels when $(N_t,N_r) \in \{(2,8),(2,16)\}$.}  \vspace{-3mm}
		\label{fig:EM_SER2}
    \end{figure}
    
        Fig.~\ref{fig:EM_SER2} compares the SERs of various detection methods for the realistic distortion scenario with time-invariant channels when $(N_t,N_r) \in \{(2,8),(2,16)\}$. 
        Fig.~\ref{fig:EM_SER2} shows that the proposed data-driven method provides additional performance gain over the model-driven method, unlike the results in Fig.~\ref{fig:EM_SER}.
        The underlying reason is that nonlinear distortion caused by the PAs and low-resolution ADCs cannot be fully captured by the generalized Gaussian model adopted in the model-driven method. 
        Therefore, the performance of the model-driven method is degraded due to a model mismatch problem. 
        Nevertheless, our model-driven method still outperforms the existing detection methods for all SNR values. 
        It is also shown that the performance gap between the proposed and existing methods is larger in Fig.~\ref{fig:EM_SER2} than that observed in Fig.~\ref{fig:EM_SER}.
        This result demonstrates that the larger the distortion caused by hardware impairments, the greater the gain provided by the proposed methods.

        \begin{figure}[t]
            \centering 
                {\epsfig{file=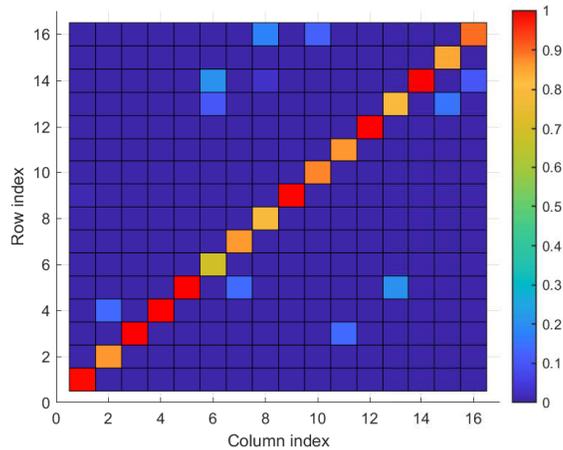, width=8.5cm}} \vspace{-3mm}
            \caption{Visualization of the entries of a transition matrix ${\bm \Theta}$ estimated by the EMNL algorithm for the realistic distortion scenario with time-invariant channels when $(N_t,N_r) = (2,8)$ and ${\sf SNR}=0$dB.}\vspace{-3mm}
            \label{fig:Confusion}
        \end{figure}

        In Fig.~\ref{fig:Confusion}, we visualize the entries of a transition matrix ${\bm \Theta}$ estimated by the EMNL algorithm for the realistic distortion scenario with time-invariant channels when $(N_t,N_r) = (2,8)$ and ${\sf SNR}=0$dB. 
        Fig.~\ref{fig:Confusion} shows that not all the diagonal entries of the transition matrix have the value of one even after the convergence of the EMNL algorithm (i.e., $I_{\rm EMNL}=20$).
        By the definition, the value of the $k$-th diagonal entry of ${\bm \Theta}$ represents the reliability of training samples labeled with $k$. 
        Therefore, the positions of diagonal entries whose values are less than one are likely to be the labels that may have incorrect values in the training data.
        The result in Fig.~\ref{fig:Confusion} implies that the EMNL algorithm does not treat all training data samples equally. Instead, the algorithm properly estimates the transition probabilities from noisy labels to true labels and treats the samples by taking into account the estimated transition probabilities.

    \begin{figure}[t]
		\centering 
		\subfigure[MIMO systems]
		{\epsfig{file=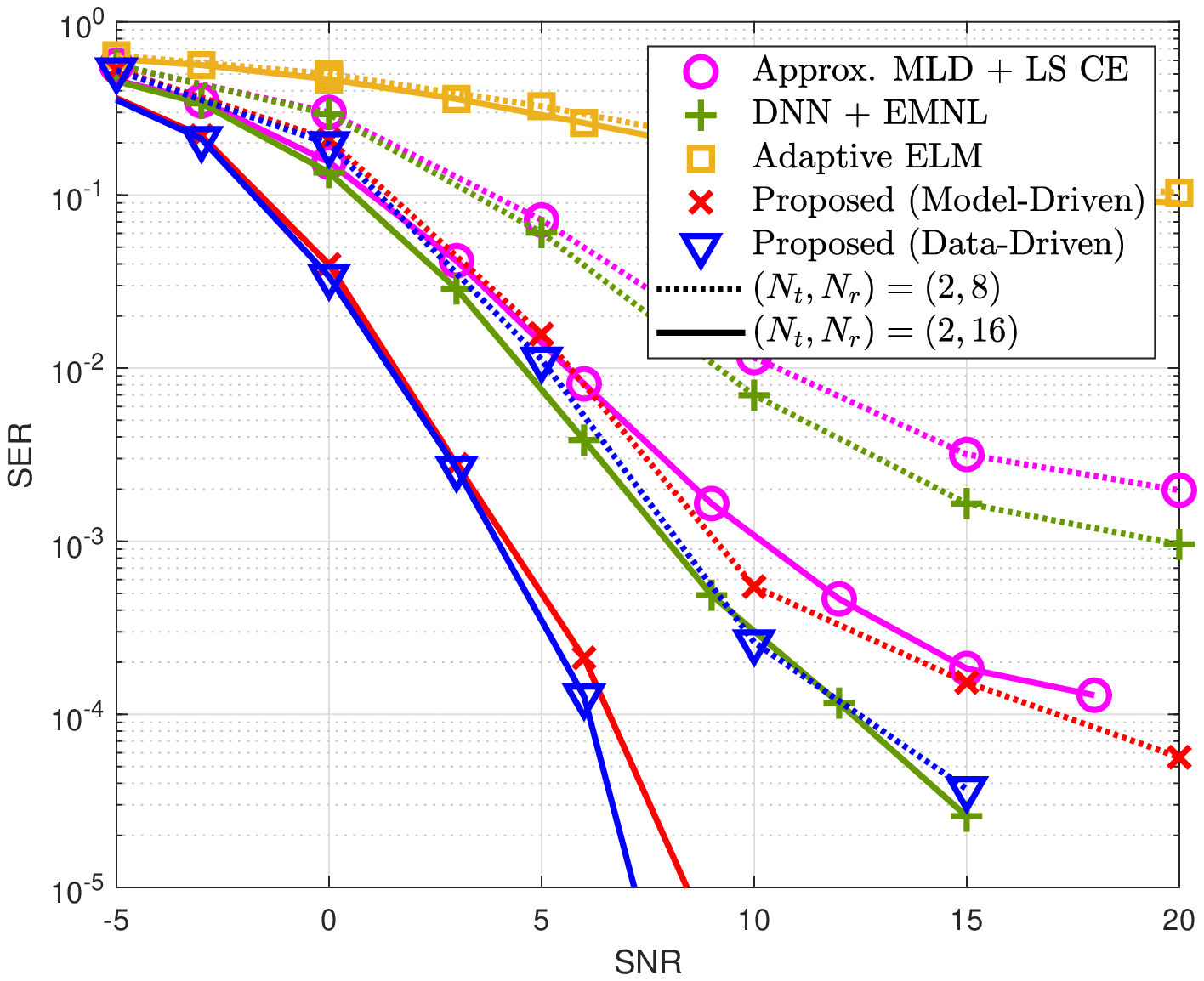, width=8cm}}
  \subfigure[SIMO systems]
		{\epsfig{file=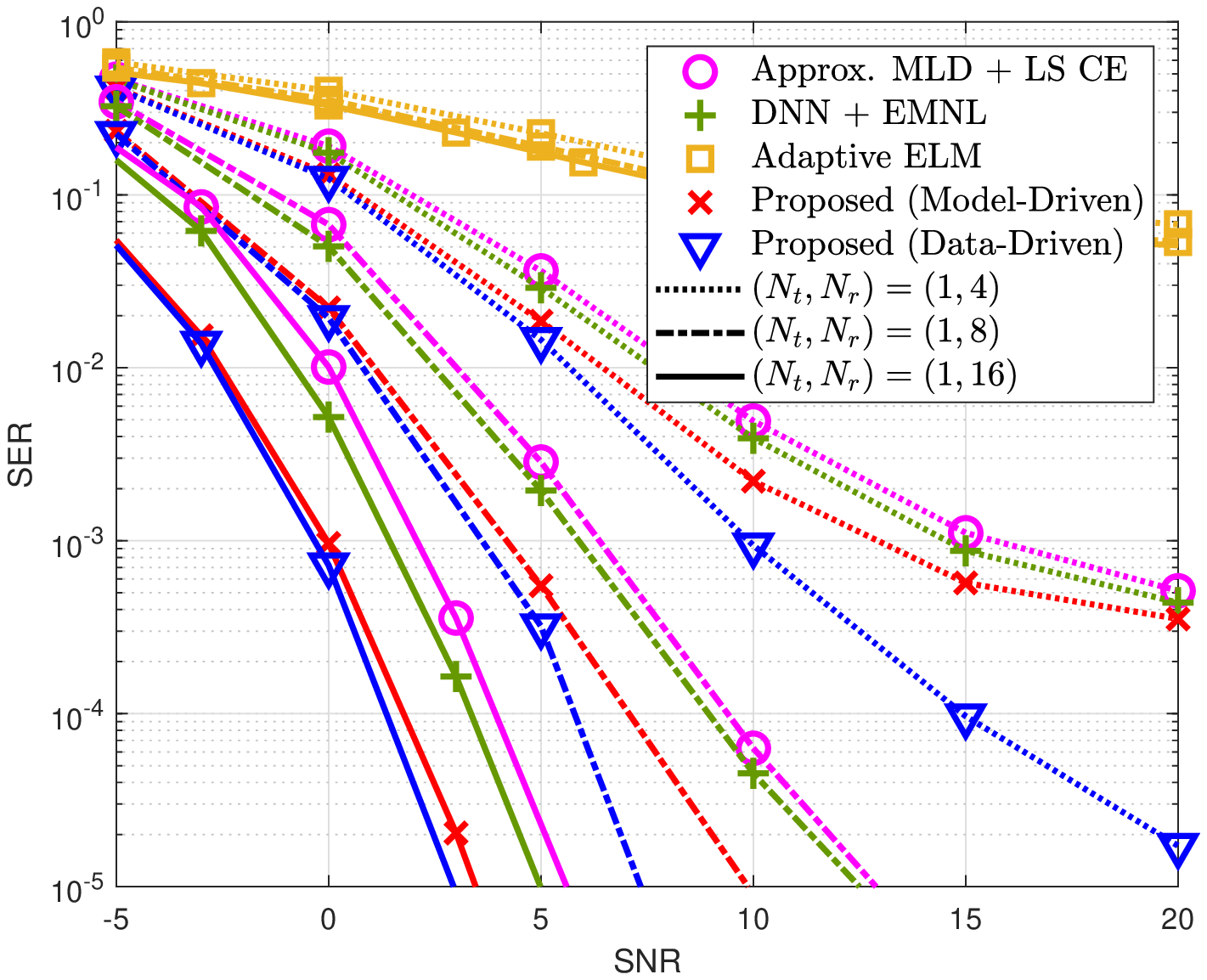, width=8cm}}
		\caption{SER comparison of various detection methods for the realistic distortion scenario with time-varying channels.}  \vspace{-3mm}
		 \label{fig:SER_Varying}
    \end{figure}

    Fig.~\ref{fig:SER_Varying} compares the SERs of various detection methods for the realistic distortion scenario with time-varying channels. 
    It should be first noted that compensating for hardware impairments in time-varying channels is more difficult due to the variations in the distortion over time.   
    Even for this challenging scenario, the proposed methods are shown to outperform the existing detection methods for all SNR values regardless of the antenna configuration. 
    In particular, for the same antenna configuration  (e.g., $(N_t,N_r)=(2,8)$ or $(N_t,N_r)=(2,16)$), the performance gap between the proposed and existing methods is larger in Fig.~\ref{fig:SER_Varying}(a) than that observed in Fig.~\ref{fig:EM_SER2}.
    This result demonstrates the robustness of the proposed methods against not only nonlinear distortion but also temporal channel variations. 
    Fig.~\ref{fig:SER_Varying} also shows that the performance gap between the model-driven and data-driven methods increases as the number of receive antennas decreases for both the single-input multiple-output (SIMO) and MIMO systems.
    In general, the noisy training data set generated using our strategy tends to have a higher number of false labels when the number of receive antennas is smaller.
    Therefore, this result demonstrates that our robust DNN training algorithm works well even with noisy training data by effectively mitigating the undesirable impacts of the false labels.

    \begin{table}[ht]
        \centering
            \caption{SER comparison between the proposed methods and a na\"ive DNN-based detection method for the realistic distortion scenario with time-varying channels.}
        \begin{tabular}[h]{lccc}
            \hline
            &Model-Driven &Data-Driven & Na\"ive DNN \\
            \hline
            $(N_t,N_r)=(1,4)$, 20\text{dB}&$3.53\times 10^{-4}$ &$1.72\times 10^{-5}$ &$1.10\times 10^{-4}$\\
            $(N_t,N_r)=(1,8)$, 7\text{dB}&$9.84\times 10^{-5}$ &$4.28\times 10^{-5}$ &$6.20\times 10^{-5}$\\
            $(N_t,N_r)=(1,16)$, 2\text{dB}&$7.08\times 10^{-5}$ &$3.46\times 10^{-5}$ &$5.88\times 10^{-5}$\\
            $(N_t,N_r)=(2,8)$, 15\text{dB}&$1.53\times 10^{-4}$ &$3.70\times 10^{-5}$ &$9.79\times 10^{-5}$\\
            $(N_t,N_r)=(2,16)$, 8\text{dB}&$4.83\times 10^{-5}$ &$1.67\times 10^{-5}$ &$2.36\times 10^{-5}$\\
            \hline
        \end{tabular}\label{table:SER_comp}
    \end{table}
    Table~\ref{table:SER_comp} compares the SERs of the proposed detection methods and a na\"ive DNN-based detection method for the realistic distortion scenario with time-varying channels.
    The na\"ive DNN-based detection method uses a conventional DNN training algorithm during the DNN training process with noisy training data $\hat{\mathcal{D}}_{\rm MD}$, instead of the robust DNN training algorithm used in the proposed data-driven detection method. 
    Table~\ref{table:SER_comp} shows that the data-driven detection method outperforms the na\"ive DNN-based method for all considered configurations. 
    This result demonstrates that our robust DNN training algorithm effectively mitigates the overfitting problem caused by noisy training data, compared to the conventional DNN training algorithm. 
    Nevertheless, the na\"ive DNN-based method still achieves a lower SER than the proposed model-driven detection method, demonstrating the validity of the DL-based approach to circumvent the model mismatch problem. 


    \section{Conclusion}\label{Sec:Conclusion}
    In this paper, we have explored a data detection problem in MIMO communication systems with hardware impairments.
    To mitigate the degradation of detection performance due to nonlinear and unknown distortion in baseband received signals, we have proposed two novel detection methods, referred to as model-driven and data-driven.
    Our model-driven method employs a generalized Gaussian distortion model to approximate the conditional distribution of the received signal.
    To address the model mismatch problem in the model-driven method, our data-driven method leverages a DNN for approximating the APP vector for the received signal. 
    We have developed robust training algorithms to accurately train the distortion model and the DNN from noisy training data generated using coarse detection outputs. 
    A prominent advantage of our methods is that they do not require additional training overhead beyond the traditional pilot overhead for channel estimation.
    Using simulations, we have demonstrated that the proposed detection methods outperform existing detection methods under various hardware impairment scenarios. 
    Our simulations also verify the robustness of the proposed methods against not only hardware impairments but also temporal channel variations.

    An important direction for future research is to develop a low-complexity detection method for MIMO systems with hardware impairments, by reducing the computational overhead of the proposed detection methods. 
    Another important research direction is to extend the proposed detection methods for application in frequency-selective MIMO systems with hardware impairments. 
    It would also be possible to improve the performance of the proposed detection methods by leveraging online learning or meta-learning techniques. 
    
\bibliographystyle{unsrt}

\begin{thebibliography}{1} 
            \bibitem{Conference} J. Kwon, Y.-S. Jeon, and H. V. Poor, ``MIMO detection under hardware impairments via learning from noisy labels,'' 
            in {\em Proc. IEEE Global Commun. Conf. (GLOBECOM),}  Dec. 2022, pp. 3192--3197. 
            
		\bibitem{5G} J. G. Andrews, S. Buzzi, W. Choi, S. V. Hanly, A. Lozano, A. C. K. Soong, and J. C. Zhang,
		``What Will 5G Be?'' 
		{\em IEEE J. Sel. Areas Commun.,} vol. 32, no. 6, pp. 1065--1082, June 2014.
		
		\bibitem{6G} W. Saad, M. Bennis, and M. Chen,
		``A vision of 6G wireless systems: Applications, trends, technologies, and open research problems,'' 
		{\em IEEE Network}, vol. 34, no. 3, pp. 134--142, May/June 2020.
		
        
        \bibitem{lowADC : 3} O. Orhan, E. Erkip, and S. Rangan, ``Low power analog-to-digital conversion in millimeter wave systems: Impact of resolution and bandwidth on performance,''
        in {\em Proc. Inf. Theory Applications Workshop (ITA),} 
        Feb. 2015, pp. 191--198.
        
	\bibitem{lowADC : Magazine} Y.-S. Jeon, D. Kim, S.-N. Hong, N. Lee, and R. W. Heath, 
        ``Artificial intelligence for physical-layer design of MIMO communications with one-bit ADCs,'' 
        {\em IEEE Commun. Mag.,} vol. 60, no. 7, pp. 76--81, July 2022.
        
        \bibitem{HI : PA 1}  H. Wang, F. Wang, H. T. Nguyen, S. Li, T. Y. Huang, A. S. Ahmed, M. E. D. Smith, N. S. Mannem, and J. Lee, 
        ``Power amplifiers performance survey 2000-present,'' 
        [Online]. Available: https://gems.ece.gatech.edu/PA\_survey.html
        
        \bibitem{HI:THz} T. Mao, Q. Wang, and Z. Wang, 
        ``Receiver design for the low-cost TeraHertz communication system with hardware impairment,'' 
        in {\em Proc. IEEE Int. Conf. Commun. (ICC),} Jun. 2020, pp. 1--6.
            
        \bibitem{HI : PA 2} E. Björnson, J. Hoydis, M. Kountouris, and M. Debbah,
        ``Massive MIMO systems with non-ideal hardware: Energy efficiency, estimation, and capacity limits,'' 
        {\em IEEE Trans. Inf. Theory,} vol. 60, no. 11, pp. 7112--7139, Nov. 2014.
      		
		
        \bibitem{HI : book} T. Schenk, 
        {\em RF Imperfections in High–Rate Wireless Systems,} 
        New York, NY, USA: Springer, 2008.
        
        \bibitem{HI : mmWave} M. Wu, D. Wuebben, A. Dekorsy, P. Baracca, V. Braun, and H. Halbauer,
        ``Hardware impairments in millimeter wave communications using OFDM and SC-FDE,''  
        in {\em Proc. ITG Workshop Smart Antennas (WSA),} Mar. 2016, pp. 1--8.

        \bibitem{HI : PA 3} C. Studer, M. Wenk, and A. Burg,
        ``MIMO transmission with residual transmit-RF impairments,'' 
        in {\em Proc. ITG Workshop Smart Antennas (WSA),} Feb. 2010, pp. 189--196.
  

        \bibitem{Learning(Model) : 1} M. Abdelaziz, L. Anttila, A. Brihuega, F. Tufvesson, and M. Valkama,  
        ``Digital predistortion for hybrid MIMO transmitters,'' 
        {\em IEEE J. Sel. Topics Signal Process.,} vol. 12, no. 3, pp. 445--454, Jun. 2018.

        \bibitem{Learning(Model) : 2} S. Lee, M. Kim, Y. Sirl, E.-R. Jeong, S. Hong, S. Kim, and Y. H. Lee,    
        ``Digital predistortion for power amplifiers in hybrid MIMO systems with antenna subarrays,'' 
        in {\em Proc. IEEE 81st Veh. Technol. Conf. (VTC),} May 2015, pp. 1--5.

        \bibitem{Learning(Model) : 3}  M. Abdelaziz, L. Anttila, and M. Valkama, 
        ``Reduced-complexity digital predistortion for massive MIMO,'' 
        in {\em Proc. IEEE Int. Conf. Acoust. Speech Signal Process. (ICASSP),} Mar. 2017, pp. 6478--6482.

        \bibitem{Learning DNN : 3 (Predistortion)} Y. Wu, U. Gustavsson, A. G. i. Amat, and H. Wymeersch,
        ``Residual neural networks for digital predistortion,'' 
        in {\em Proc. IEEE Global Commun. Conf. (GLOBECOM),} Dec. 2020, pp. 1--6.

        \bibitem{Learning DNN : 4 (Post compensation)} H. Liu, X. Yang, P. Chen, M. Sun, B. Li, and C. Zhao,
        ``Deep learning based nonlinear signal detection in millimeter-wave communications,'' 
        {\em IEEE Access,} vol. 8, pp. 158883--158892, Sep. 2020.




        \bibitem{DNN Application : 1 (DetNet)} N. Samuel, T. Diskin, and A. Wiesel,
        ``Learning to detect,'' 
        {\em IEEE Trans. Signal Process.,}  vol. 67, no. 10, pp. 2554--2564, Feb. 2019.

        \bibitem{DNN Application : 2 (ViterbiNet)} N. Shlezinger, N. Farsad, Y. C. Eldar, and A. J. Goldsmith,
        ``ViterbiNet: A deep learning based viterbi algorithm for symbol detection,'' 
        {\em IEEE Trans. Wireless Commun.,} vol. 19, no. 5, pp. 3319--3331, May 2020.

        \bibitem{DNN Application : 3 (OAMPNet)} H. He, C.-K. Wen, S. Jin, and G. Y. Li,
        ``Model-driven deep learning for MIMO detection,'' 
        {\em IEEE Trans. Signal Process.,} vol. 68, pp. 1702--1715, Feb. 2020.


        \bibitem{Learning DNN : 1 (Naive)} Q. Hu, F. Gao, H. Zhang, G. Y. Li, and Z. Xu,
        ``Understanding deep MIMO detection,'' 
        arXiv:2105.05044 [eess.SP], May 2021. [Online]. Available: https://arxiv.org/abs/2105.05044
  
        \bibitem{Learning DNN : 5 (Lord-Net)} S. Khobahi, N. Shlezinger, M. Soltanalian, and Y. C. Eldar, 
        ``LoRD-Net: Unfolded deep detection network with low-resolution receivers,'' 
        {\em IEEE Trans. Signal Process.,} vol. 69, pp. 5651--5664, Oct. 2021.

        \bibitem{Learning DNN : 2 (MP)} D. Gao, Q. Guo, G. Liao, Y. C. Eldar, Y. Li, Y. Yu, and B. Vucetic,
        ``Signal detection in MIMO systems with hardware imperfections: Message passing on neural networks,'' 
        arXiv:2210.03911 [eess.SP], Oct. 2022. [Online]. Available: https://arxiv.org/abs/2210.03911

  
        \bibitem{Learning : 1 (SLD)} Y.-S. Jeon, S.-N. Hong, and N. Lee,
        ``Supervised-learning-aided communication framework for MIMO systems with low-resolution ADCs,'' 
        {\em IEEE Trans. Veh. Tech.,} vol. 67, no. 8, pp. 7299--7313, Aug. 2018.
        
        \bibitem{Learning : 2 (SLD)} S. Kim and S.-N. Hong,
        ``A supervised-learning detector for multihop distributed reception systems,'' 
        {\em IEEE Trans. Veh. Tech.,} vol. 68, no. 2, pp. 1958--1962, Feb. 2019.	
		
        \bibitem{Learning : 3 (ELM)} D. Gao, Q. Guo, and Y. C. Eldar,
        ``Massive MIMO as an extreme learning machine,'' 
        {\em IEEE Trans. Veh. Tech.,} vol. 70, no. 1, pp. 1046--1050, Jan. 2021.
       
        \bibitem{NLNN} A. J. Bekker and J. Goldberger,
        ``Training deep neural-networks based on unreliable labels,'' 
        in {\em Proc. IEEE Int. Conf. Acoustics, Speech Signal Process. (ICASSP),} Mar. 2016, pp. 2682--2686.
  
        \bibitem{Deep : SELFIE} H. Song, M. Kim, D. Park, Y. Shin, and J. G. Lee,
        ``Selfie: Refurbishing unclean samples for robust deep learning,''  
        in {\em Proc. Int. Conf. Mach. Learn. (ICML),} June 2019, pp. 5907--5915.

        \bibitem{Deep : Self-adaptive} L. Huang, C. Zhang, and H. Zhang,
        ``Self-adaptive training: Beyond empirical risk minimization,''  
        in {\em Adv. Neural Inf. Process. Syst. (NeurIPS),} Dec. 2020, pp. 19365--19376.
        

  


        \bibitem{Deep : Memorization1} D. Arpit, S. Jastrz\c{e}bski, N. Ballas, D. Krueger, E. Bengio, M. S. Kanwal, T. Maharaj, A. Fischer, A. Courville, Y. Bengio, and S. Lacoste-Julien,
        ``A closer look at memorization in deep networks,''  
        in {\em Proc. Int. Conf. Mach. Learn. (ICML),} Jun. 2017, pp. 233--242.
        
        \bibitem{Deep : Fool} A. Nguyen, J. Yosinski, and J. Clune,
        ``Deep neural networks are easily fooled: High confidence predictions for unrecognizable images,'' 
        in {\em Proc. IEEE Conf. Comput. Vis. Pattern Recognit. (CVPR),} Jun. 2015, pp. 427--436.

        
        \bibitem{Deep : Generalization} C. Zhang, S. Bengio, M. Hardt, B. Recht, and O. Vinyals,
        ``Understanding deep learning (still) requires rethinking generalization,''  
        {\em Communications of the ACM,} vol. 64, no. 3, pp. 107--115, Feb. 2021.





        \bibitem{Deep : Coteaching} B. Han, Q. Yao, X. Yu, G. Niu, M. Xu, W. Hu, I. Tsang, and M. Sugiyama,
        ``Co-teaching: Robust training of deep neural networks with extremely noisy labels,''  
        in {\em Adv. Neural Inf. Process. Syst. (NeurIPS),} Dec. 2018, pp. 8536--8546.
    
        \bibitem{Deep : MentorNet} L. Jiang, Z. Zhou, T. Leung, L.-J. Li, and L. Fei-Fei,
        ``MentorNet: Learning data-driven curriculum for very deep neural networks on corrupted labels,''  
        in {\em Proc. Int. Conf. Mach. Learn. (ICML),} July 2018, pp. 2303--2313.
 
        \bibitem{Deep : Coregularization} H. Wei, L. Feng, X. Chen, and B. An,
        ``Combating noisy labels by agreement: A joint training method with co-regularization,''  
        in {\em Proc. IEEE Conf. Comput. Vis. Pattern Recognit. (CVPR),} June 2020, pp. 13726--13735.
    
        \bibitem{Deep : Disagreement} X. Yu, B. Han, J. Yao, G. Niu, I. Tsang, and M. Sugiyama,
        ``How does disagreement help generalization against label corruption?,''  
        in {\em Proc. Int. Conf. Mach. Learn. (ICML),} July 2019, pp. 7164--7173.
        

        \bibitem{Deep : Data uncertainity} A. Kendall and Y. Gal,
        ``What uncertainties do we need in bayesian deep learning for computer vision?,''  
        in {\em Adv. Neural Inf. Process. Syst. (NeurIPS),} Mar. 2017, pp. 5574--5584.

        \bibitem{Deep : Bootstraping} S. Reed, H. Lee, D. Anguelov, C. Szegedy, D. Erhan, and A. Rabinovich,
        ``Training deep neural networks on noisy labels with bootstrapping,''  
        arXiv:1412.6596 [cs.CV], Apr. 2015.
        [Online]. Available: https://arxiv.org/abs/1412.6596
        
        \bibitem{PA : 1} A. A. M. Saleh,
        ``Frequency-independent and frequency-dependent nonlinear models of TWT amplifiers,'' 
        {\em IEEE Trans. Commun.,}  vol. 29, no. 11, pp. 1715--1720, Nov. 1981.

  
    \end{thebibliography}
	
\end{document}